\documentclass[twocolumn,aps,prb,longbibliography]{revtex4-1} %linenumbers %superscriptaddress
\usepackage{graphicx,amssymb,amsmath,psfrag,xcolor,bm,xfrac}
\usepackage[colorlinks=true,citecolor=blue,linkcolor=blue,urlcolor=blue]{hyperref}

\def\targetformat{arXiv}

\newcommand{\para}{{\mkern3mu\vphantom{\perp}\vrule depth 0pt\mkern2mu\vrule depth 0pt\mkern3mu}}

\usepackage{framed}
\definecolor{shadecolor}{gray}{0.95}

\usepackage{amsthm}
\newtheorem*{theorem*}{Theorem}

\begin{document}
	\title{Topology and edge states survive quantum criticality between topological insulators}
	\author{Ruben Verresen}
	\affiliation{Department of Physics, Harvard University, Cambridge MA 02138, USA}
	\date{\today}
	
\maketitle

\textbf{It is often thought that emergent phenomena in topological phases of matter are destroyed
when tuning to a critical point.
In particular, topologically protected edge states supposedly delocalize when the bulk correlation length diverges.
We show that this is not true in general.
Edge states of topological insulators or superconductors remain exponentially localized---despite a vanishing band gap---if the transition increases the topological index.
This applies to all classes where the topological classification is larger than $\mathbb Z_2$, notably
including Chern insulators. Moreover, these edge states are stable to disorder, unlike in topological semi-metals.
This new phenomenon is explained by generalizing band (or mass) inversion---a unifying perspective on topological insulators---to kinetic inversion.
In the spirit of the bulk-boundary correspondence, we also identify topological invariants at criticality, which take half-integer values and separate topologically-distinct universality classes by a multi-critical point.
This work enlarges the scope of topological protection and stability by showing that bulk energy gaps can be unnecessary.
Experimental probes and stability to interactions are discussed.}

\begin{figure}
	\includegraphics[scale=1,trim=0.1cm 0.15cm 0.15cm 0.1cm,clip]{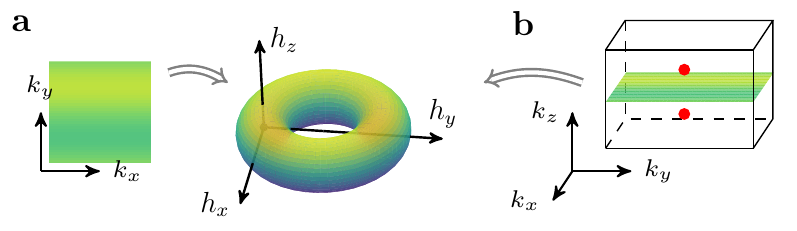}
	\caption{\textbf{Conventional topological band insulators and metals rely on energy gaps.} \textbf{a}, Topological insulators map momentum space (here, two-dimensional) into the space of single-particle Hamiltonians (here, three-dimensional). The number of times this surface encloses the origin is a topological invariant (here, the Chern number $\mathcal C =1$). A nonzero bulk energy gap implies that the surface cannot intersect the origin. \textbf{b}, A topological Weyl semi-metal. The bulk energy gap vanishes at the two solid red points, but lower-dimensional momentum slices avoiding these points have an effective nonzero gap and well-defined Chern number.
		\label{fig:topology} }
\end{figure}

Topological phases of matter are greater than the sum of their parts. Undetectable by any local order parameter, their non-triviality is encoded in global topological invariants. This is well-understood in the insulating case, i.e., when there is an energy threshold to creating excitations above the ground state---referred to as an energy or band gap. This is beautifully illustrated in non-interacting fermionic lattice models, e.g., a two-band Hamiltonian in momentum space, $\mathcal H(\bm k)= \bm{h}(\bm k) \bm{\cdot \sigma}$ where $\bm \sigma = (\sigma_x,\sigma_y,\sigma_z)$ are the Pauli matrices. In, say, two spatial dimensions, $\bm h ( \bm k)$ is a closed surface as shown in Fig.~\hyperref[fig:topology]{1a}, and its minimum distance to the origin gives the energy gap. Hence, a nonzero gap implies that the number of times this surface wraps around the origin is a well-defined topological invariant---in this case known as the Chern number $\mathcal C$. Physically, this integer manifests itself in a quantized anomalous Hall conductance $\sigma_{xy} = \frac{e^2}{h} \mathcal C$ and $|\mathcal C|$ perfectly-conducting edge modes \cite{Laughlin81,Halperin82,Haldane88}. 

Topological phenomena are much less understood when the energy gap vanishes, such as at quantum critical points\cite{Sachdev01}.
According to folklore, gapless systems \emph{cannot} host quantized invariants or localized edge modes.
The penetration depth of edge modes is often thought to coincide with the bulk correlation length, and the latter blows up with the inverse gap as one approaches criticality.
Even topological semi-metals comply with this intuition. For instance, while a 3D Weyl semi-metal\cite{Turnerbook13} is gapless at particular points in momentum space, any 2D slice avoiding these points is a gapped lower-dimensional system (see Fig.~\hyperref[fig:topology]{1b}).
The Chern number of the latter protects the famous Fermi arcs. Topological semi-metals thus
rely on a \emph{momentum-dependent} energy gap, meaning that disorder---which connects momenta---destabilizes them.

In recent years, a plethora of models have come to light challenging the conventional viewpoint\cite{Kestner11,Cheng11,Fidkowski11longrange,Sau11,Grover12,Kraus13,Keselman15,Iemini15,Lang15,Montorsi17,Ruhman17,Scaffidi17,Jiang18,Zhang18,Verresen18,Parker18,Keselman18,Verresen19}.
They are gapless yet host localized topologically-protected edge modes---without needing momentum conservation (see Ref.~\onlinecite{Verresen19} for a unifying perspective). However, most of these works are limited to 1D systems---rare exceptions\cite{Grover12,Scaffidi17} rely on \emph{additional} gapped degrees of freedom.
Here, we are interested in the interplay between topology and criticality without \emph{any} energy gap.

This work shows how topological invariants and localized edge modes can exist at phase transitions between topological insulators\cite{Hasan10} in general dimension. 
For instance, while the Chern insulator transitions between $\mathcal C=0 \leftrightarrow \mathcal C=1$ and between $\mathcal C=1 \leftrightarrow \mathcal C=2$ are both described by a single Dirac cone, we generalize $\mathcal C$ to criticality, with the former having $\mathcal C = \frac{1}{2}$ and the latter $\mathcal C= \frac{3}{2}$, making them topologically distinct Dirac cones---separated by a phase transition! Moreover, for $\mathcal C= \frac{3}{2}$ there is an exponentially localized chiral edge mode. We emphasize that, although critical points are by definition fine-tuned, the phenomena we discuss are not; the edge mode is stable to gap-opening. Said in reverse, the edge mode of a Chern insulator with $\mathcal C=1$ remains localized whilst tuning to $\mathcal C=2$, instead of smearing out with the diverging bulk correlation length.
We first introduce these invariants, after which we discuss the edge modes. %\linebreak

$\;$ 

%\newpage

\onecolumngrid

\begin{figure}[h]
	\includegraphics[scale=1.02]{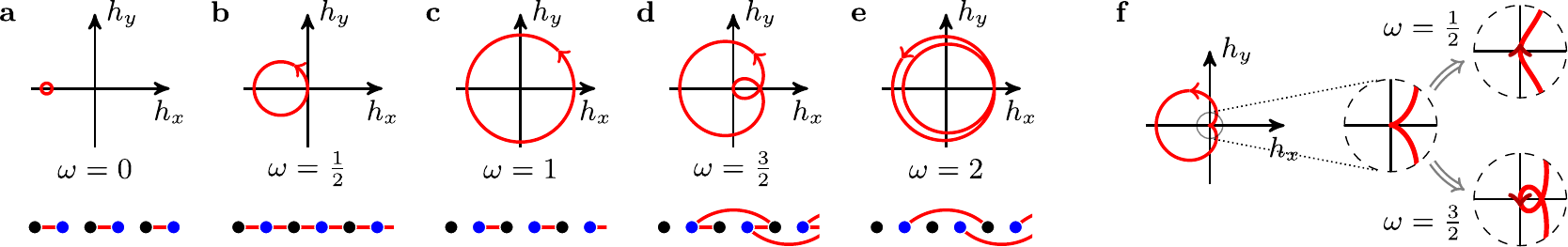}
	\caption{
		\textbf{Winding numbers
		%and edge modes
		beyond the gapped paradigm.} \textbf{a--e}, Gapped and gapless one-dimensional topological wires with integer and half-integer topological invariant $\omega$, respectively. The single-particle Hamiltonian $\mathcal H(k) = \bm{h}(k) \bm{ \cdot \sigma}$ (top) and real-space hopping amplitudes (bottom) are shown.
		%, with localized edge modes if $|\omega| > \frac{1}{2}$. 
		The gapless wire with $\omega = \frac{3}{2}$ is topologically non-trivial, resulting in a localized edge mode (see Fig.~\ref{fig:domain}). \textbf{f}, The (cardioid) mother theory which can be perturbed into the five phases shown in \textbf{a--e}. A perturbation linear in $k$ (near $k \approx 0$) flows to $\omega=\frac{1}{2}$ or $\omega=\frac{3}{2}$. %, depending on the choice of sign.
		It is a topological transition between two gapless phases.
		\label{fig:winding}}
\end{figure}
\twocolumngrid

%\noindent

\textbf{Topological invariants.} To define and understand topological invariants at criticality, we first consider a 1D setting. Symmetry is then necessary to have gapped topological phases labeled by integers (similar to $\mathcal C \in \mathbb Z$ above). We can consider, e.g., topological insulators protected by sublattice symmetry (`class AIII') or topological superconductors protected by spinless time-reversal symmetry (`class BDI')\cite{Kitaev09,Schnyder08,Ryu10}. Both classes have a unified description in the single-particle Hamiltonian: the symmetry pins $h_z(k) = 0$. We can thus visualize $\bm h(k)$ as a loop in the plane, its winding number around the origin being a topological invariant $\omega = \frac{1}{2\pi} \int \mathrm d \left( \textrm{arg}(\bm h) \right) \in \mathbb Z$. Figures~\hyperref[fig:winding]{2a,c,e} show examples with winding $\omega=0,1,2$.

The transitions between these phases are also shown, where the gap closes at $\bm h(0)=0$.
The two loops in Fig.~\hyperref[fig:winding]{2b,d} are similar for $k \approx 0$: $\bm h$ is linear, implying that the low-energy theory is a relativistic fermion. 
However, Fig.~\hyperref[fig:winding]{2d} shows that at larger momenta, the transition between $\omega=1\leftrightarrow \omega = 2$ encircles the origin.
To capture this `high-energy' topological information, we introduce an invariant calculating the winding whilst omitting an infinitesimal neighborhood of the gapless point:
\begin{equation}
\omega = \frac{1}{2\pi} \lim_{\varepsilon \to 0} \int_{|\bm h| > \varepsilon}
\mathrm d \left( \textrm{arg}(\bm h)\right) 
\quad \in \frac{1}{2} \mathbb Z. \label{eq:defomega}
\end{equation}
This equals the average of $\omega$ over the neighboring gapped phases, as one can show by gapping the omitted low-energy part (see \hyperref[sec:methods]{Methods}; this only relies on the leading non-zero derivative at the gapless point being continuous, which is a natural generalization of the usual continuity condition on $\bm h(k)$ and corresponds to enforcing locality in real space).
For the two critical points above, we thus read off $\omega=\frac{1}{2}$ and $\omega=\frac{3}{2}$, respectively.
To gain some intuition: this is a half-integer since a line through the origin spans a $180^\circ$ angle.
Hence, as long as the derivative $\partial_k \bm h(0)$ is nonzero, we cannot change $\omega$, making it a robust property of the low-energy universality class.
The only way we can connect these two critical points is by tuning through a multi-critical point which is no longer described by a (single) relativistic fermion. A possible
`topological phase transition of phase transitions' is shown in Fig.~\hyperref[fig:winding]{2f}, with a quadratic dispersion ($|\bm h| \sim k^2$), where the linear term changes sign.

We thus have transitions which are locally described by the same universality class yet non-locally distinguished by topology. While 1D cases have been studied before with other methods\cite{Motrunich01,Verresen18,Jones19}, Eq.~\eqref{eq:defomega} generalizes to arbitrary dimensions and symmetry classes: one takes the definition of a gapped topological invariant and cuts out an infinitesimal region for each gapless point.
This equals the average of the invariant over the neighboring gapped phases.
To illustrate this in 2D, we generalize the Chern number to any case where the low-energy theory is local (i.e., $\bm h(\bm k)$ is smooth wherever it vanishes):
\begin{equation}
\mathcal C = \frac{1}{4\pi} \lim_{\varepsilon \to 0} \iint_{|\bm h (\bm k)|>\varepsilon} \bm{\hat h \cdot} \big( \partial_{k_x} \bm{\hat h} \times \partial_{k_y} \bm{\hat h} \big)
\mathrm d^2 \bm k \quad \in \frac{1}{2} \mathbb Z, \label{eq:defchern}
\end{equation}
where $\bm{\hat h} = \frac{\bm h}{|\bm h|}$.
The transition between $\mathcal C= n$ and $\mathcal C= n+1$ is then a Dirac cone with $\mathcal C = n + \frac{1}{2}$. This agrees with the parity anomaly of $(2+1)$-dimensional quantum electrodynamics, requiring a Chern-Simons term with half-integer prefactor, or more precisely, an $\eta$-term\cite{Redlich84,Alvarez85,Haldane88,Seiberg16} (i.e., the action will have a topological term $S_\textrm{top} = \mathcal C \pi \eta(A)$).
What was perhaps not appreciated is that this allows for topologically distinct Dirac cones separated by multi-criticality.
A possible transition between, say, $\mathcal C = \frac{1}{2}$ and $\mathcal C = \frac{3}{2}$ would have the Dirac cone become quadratic along $k_x$; alternatively, a second Dirac cone could appear.

\textbf{Edge modes.} Having established bulk topological invariants, we turn to one of their most fascinating consequences: topologically-protected states localized near the system's boundary. For gapped topological phases, this follows from the existence of bulk phase transitions.
Indeed, the boundary of a topological phase can be interpreted as a spatial interface between the topological phase and the trivial ``vacuum''. The phase transition is thus spatially sandwiched between the two, leading to a zero-energy mode. We generalize this concept to the critical case, giving rise to exponentially-localized edge modes, despite the absence of gapped degrees of freedom.

\begin{figure}
	\includegraphics[scale=1,trim=0.10cm .2cm 0.04cm 0.12cm,clip]{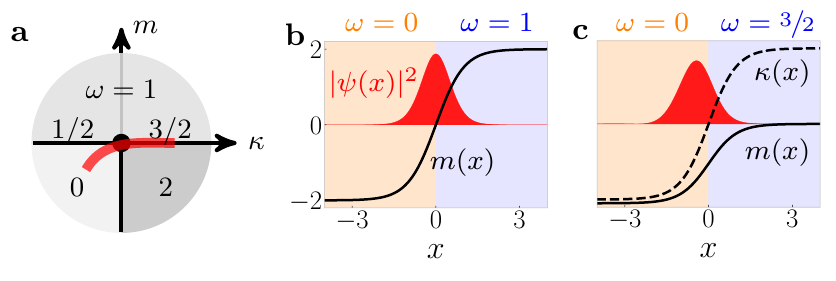}
	\caption{\textbf{Zero-energy edge modes.} \textbf{a}, Universal phase diagram near the multi-critical point in Fig.~\hyperref[fig:winding]{2f}, showing the values of the topological invariant. The critical line with $\omega=\frac{3}{2}$ is not perturbatively close to the trivial phase ($\omega=0$). Any path between the two (e.g., red line) tunes through a singular point (e.g., the multi-critical point).
	\textbf{b}, A spatial interface between the trivial and topological gapped phase. This hosts an exponentially-localized zero-energy mode (red) where the mass term changes sign.
	\textbf{c}, A spatial interface corresponding to the path in \textbf{a}. There is again a localized mode, now due to the kinetic term changing sign (related to passing through, or around, the multi-critical point). \label{fig:domain} }
\end{figure}

As we did for the topological invariants, we first explain the edge modes in the 1D setting.
The claim is that the phase transition from $\omega = n$ to $\omega = n+1$ (for which $\omega = n + \frac{1}{2}$) has $n\geq 0$ exponentially-localized edge modes, which we illustrate for $n=1$. The key is that $\omega= \frac{3}{2}$ cannot be perturbed into the trivial phase. A typical phase diagram is shown in Fig.~\hyperref[fig:domain]{3a}: the red path interpolating between $\omega=\frac{3}{2}$ and $\omega=0$ passes through the multi-critical point. (Even if one circumvents this by passing through the gapped regions where $\omega=1$ or $2$, there is still a transition to $\omega=0$.) This inevitability of a singularity between $\omega = 0$ and $\omega=\frac{3}{2}$ implies an edge mode at their spatial interface---similar to the gapped case---as we argue now.

To make this precise, let us review the argument for the familiar gapped case\cite{Jackiw76}.
To set up a spatial interface between $\omega=0$ and $\omega=1$, we need a model realizing both phases.
For this, simply move the loop in Fig.~\hyperref[fig:winding]{2b} to the left or right: $h(k) = h_x + i h_y = \left( e^{ik} -1 \right) + m$,
which is trivial for $m<0$.
To focus on universal low-energy physics, we zoom in near the gap-closing at $k\approx 0$, giving $h(k) \approx ik + m$.
Transforming to real space, we arrive at $D_x := h(-i\partial_x) = \partial_x + m$, or more completely:
\begin{equation}
\mathcal H(x) = \left( \begin{array}{cc} 0 & D_x^\dagger \\ D_x & 0\end{array} \right)
\textrm{ has }
\omega = \left\{ \begin{array}{ll}
0 & \textrm{if } m <0,\\
1 & \textrm{if } m >0.
\end{array} \right. \label{eq:gapped}
\end{equation}
We can now consider a mass profile $m(x)$ interpolating from $\omega=0$ to $\omega=1$, as shown in Fig.~\hyperref[fig:domain]{3b}. Jackiw and Rebbi\cite{Jackiw76} realized that this supports a localized mode.
Indeed, the zero-energy condition on {\footnotesize $\left( \begin{array}{c} \psi(x) \\ 0 \end{array} \right)$} is
\begin{equation}
D_x \psi(x) = 0\quad \Rightarrow \quad \psi(x) \propto e^{-\int_0^x m(x')}. \label{eq:jackiwrebbi}
\end{equation}
Due to the asymptotic values $\lim_{x\to \pm \infty}m(x) = \pm |m_\infty|$ in Fig.~\hyperref[fig:domain]{3b}, this mode is exponentially localized at the interface with the inverse gap $\sim 1/|m_\infty|$. This phenomenon of mass (or band) inversion is at the root of the bulk-boundary correspondence in all topological insulators and superconductors\cite{Jeffrey10}.

The generalization to the critical case is strikingly simple yet leads to surprising results. As before, to set up an interface between $\omega=0$ and $\omega =\frac{3}{2}$, we need a model that realizes both. The
multi-critical point separating $\omega=\frac{1}{2}$ and $\omega=\frac{3}{2}$ is a natural candidate,
where $h(k) \approx k^2$.
Including linear and constant perturbations, we have $h(k) = k^2 - i \kappa k + m$, or in real space:
\begin{equation}
D_x := h(-i\partial_x) = -\partial_x^2 - \kappa(x) \partial_x + m(x). \label{eq:criticalinterface}
\end{equation}
Its phase diagram is shown in Fig.~\hyperref[fig:domain]{3a}, of which we have already encountered two limiting cases:
$m=0$ corresponds to the transition sketched in Fig.~\hyperref[fig:winding]{2f}, whereas
$\kappa \to -\infty$ (keeping $m/\kappa$ finite)
recovers Eq.~\eqref{eq:gapped}. We refer to $\kappa$ and $m$ as the kinetic and mass parameters, respectively.

We can now consider a spatial interface between the topological critical phase $\omega=\frac{3}{2}$ and the trivial vacuum, shown in Fig.~\hyperref[fig:domain]{3a,c}. Crucially, it is now not the \emph{mass} term that changes sign, but the \emph{kinetic} term.
This is sufficient to guarantee an exponentially-localized solution of the zero-energy equation $D_x \psi(x) = 0$!
Indeed, for $x \to +\infty$ (where $m \to 0$ and $\kappa \to$ constant; see Fig.~\hyperref[fig:domain]{3c}) we have $D_x = -\partial^2 - \kappa \partial = - \partial(\partial+\kappa)$, hence one solution is
constant but the other decays:
\begin{equation}
\psi(x) \sim \left\{ \begin{array}{ll}
\exp\left( - \kappa x \right) & \textrm{as } x \to + \infty, \\
\exp \left( - \frac{\kappa x}{2} \left[ 1 \pm \sqrt{ 1 + \frac{4m}{\kappa^2} } \right]  \right) & \textrm{as } x \to - \infty. \\
\end{array} \right. \label{eq:psicrit}
\end{equation}
This is exponentially localized when $\kappa$ is asymptotically positive on the right and negative on the left (since the expression in the square brackets has positive real part for $m<0$).
We call this novel phenomenon kinetic inversion.
Any microscopic information---e.g., whether we tune \emph{through} or \emph{around} the multi-critical point---is redundant, exemplifying stability.
A numerical solution is shown in Fig.~\hyperref[fig:domain]{3c}.
The edge mode is stable to arbitrary deformations of $D_x$ (i.e., higher-order derivatives) which do not cause bulk phase transitions; see \hyperref[sec:methods]{Methods} or Appendix~\ref{app:indextheorem} for a proof based on counting roots of the characteristic polynomials of $\lim_{x \to \pm \infty} D_x$.
This can be seen as an index theorem\cite{Atiyah68,Fukaya17} generalized to gapless cases.

What sets the localization length of an edge mode in a gapless system?
The above shows that the mode decays as $\exp(-\kappa x)$,
which is unrelated to any mass scale.
To see its true meaning, note that for $x \to +\infty$, Eq.~\eqref{eq:criticalinterface} becomes $D_x \propto -\partial_x - \frac{1}{\kappa} \partial_x^2$. The localization length $\sim 1/\kappa$ is thus an irrelevant perturbation---in the renormalization group (RG) sense---compared to the linear kinetic term. This is arguably the general point of view, since even in the gapped case (i.e., Eqs.~\eqref{eq:gapped} and \eqref{eq:jackiwrebbi}), the localization length $\sim 1/m$ is the prefactor of the (RG-irrelevant) kinetic term relative to the massive fixed point. In both the gapped and gapless cases, the edge mode becomes exactly localized in the RG fixed point limit.

Similar to the half-integer topological invariants, the discussion of edge modes generalizes to all dimensions.
For instance, the above carries over in its entirety to the 2D Hamiltonian $\mathcal H(x,y) = ${\scriptsize$\left( \begin{array}{cc}
-i\partial_y & D_x^\dagger \\
D_x & i \partial_y
\end{array} \right)$}.
The model in Eq.~\eqref{eq:criticalinterface} then realizes Chern insulators with Chern numbers displayed in Fig.~\hyperref[fig:domain]{3a}.
The spatial interface between $\mathcal C = 0$ and $\mathcal C = \frac{3}{2}$ hosts a mode {\scriptsize $\left( \begin{array}{c} \psi(x)e^{i k_y y} \\ 0 \end{array} \right)$} with $\psi(x)$ given by Eq.~\eqref{eq:psicrit}. This now has energy $k_y$, i.e., it is an exponentially-localized chiral mode.
We confirmed analytically and numerically that additional couplings do not delocalize the edge mode, as detailed in
Appendix~\ref{app:Chernfield}.
%the Supplemental Materials.

\begin{figure}
	\includegraphics[scale=1,trim=0.1cm 0.1cm 0.2cm 0.1cm, clip]{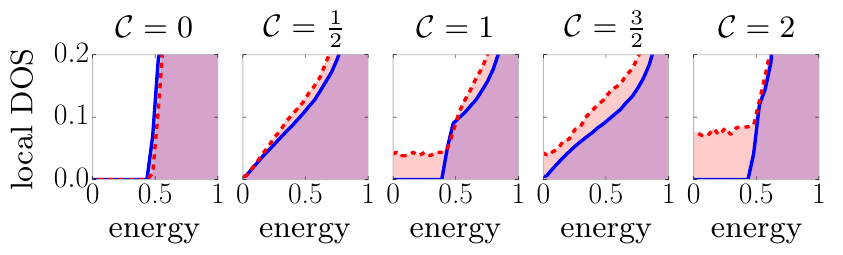}
	\caption{\textbf{Experimental signature of edge modes at critical Chern insulators.} The local density of states in the bulk (solid blue) and at the boundary (dashed red) in a lattice model for Chern insulators ($\mathcal C=0,1,2$) and their transitions ($\mathcal C=\frac{1}{2},\frac{3}{2}$). At low energies, the linear vanishing in the bulk is indicative of a Dirac cone, whereas its finite value at the boundary probes the localized edge mode. This confirms that $\mathcal C = \frac{3}{2}$ is a topology-enriched Dirac cone. \label{fig:exp}}
\end{figure}

\textbf{Experiment.} Edge modes at criticality can be experimentally diagnosed through scanning tunneling microscopy,
which measures the local density of states. Figure~\hyperref[fig:exp]{4} shows that this has a finite value at zero energy for $\mathcal C = \frac{3}{2}$ (see \hyperref[sec:methods]{Methods} for details on the lattice model).
%It would be interesting to explore a possible link with the half-integer Hall conductance at integer quantum Hall transitions\cite{Levine83,Khmelnitskii83}.
Our analysis applies to all topological insulators and superconductors which have more than one non-trivial phase, such as the $\mathbb Z$-indexed AIII, CI and DIII classes in three dimensions\cite{Schnyder08,Kitaev09}.

\textbf{Interactions.} Lastly, let us mention that topology at criticality is not a free-fermion artifact.
One way of generalizing the topological invariants to interactions is by using the anomaly perspective. Prefactors of topological response terms can be quantized even if the bulk is gapless, as we discussed for the parity anomaly of the Dirac cone.
Edge modes can also remain stable with interactions.
We recently confirmed this both analytically and numerically for $\omega = \frac{3}{2}$, with Thorngren, Jones and Pollmann\cite{Verresen19}.
The main qualitative effect of interactions is to make the edge mode algebraically-localized. In particular, for $\omega=\frac{3}{2}$ we found a universal finite-size energy splitting $\sim 1/L^{14}$---easily dwarfed by any exponential contribution.
%We stress that algebraically-localized edge modes are non-trivial as long as they are \emph{normalizable}, implying a localization length which flows to zero under RG.
It is key that the mode is \emph{normalizable}---not whether it has an exponential or algebraic profile. A normalizable mode has a finite localization length---flowing to zero under RG---and is detectable by local probes such as in Fig.~\ref{fig:exp}.

More generally, in the interacting case, we expect localized edge modes for any topological insulator transition which cannot be perturbed into the trivial phase. This is in line with the intuition that the edge mode is a spatially-confined version of the mother theory connecting the critical and the trivial phase.
To determine the universal algebraic power of the edge mode likely requires a case-by-case analysis.
We briefly summarize how to set this up (see \hyperref[sec:methods]{Methods} for details).
It is convenient to start with an RG fixed point limit of the interface studied above. In this limit, the critical system has a hard edge whose boundary condition can be derived from the original interface.
Moreover, in this same limit, the non-interacting edge mode pinches off and can be treated as a separate degree of freedom.
The goal is then to determine the dominant RG-irrelevant interaction coupling the two---its scaling dimension sets the edge mode's localization.
This approach led to the $1/L^{14}$ mentioned above\cite{Verresen19}, and it would be very interesting to analyze higher-dimensional cases. 

\textbf{Conclusion.} This
work builds toward a framework for topological phenomena beyond the gap condition.
In the non-interacting case, we non-perturbatively established edges mode at criticality. This shows that index theorems\cite{Atiyah68,Fukaya17} can be generalized, which deserves further exploration.
Relatedly, our interface set-up shows that translation symmetry is not important, but it would be interesting to study strong disorder.
We have already given pointers toward the effect of interactions which still need to be fleshed out, with anomalies offering a general perspective.
Eventually, one could hope to include all interacting cases, even those without free-fermion limits, such as symmetry breaking and topological order\cite{Wen04}.
There can be little doubt that the interplay between topology and criticality has many more surprises in store.

\vspace*{5pt}

\section*{Acknowledgements}
I thank Zhen Bi, Dan Borgnia, Tom\'a\v{s} Bzdu\v{s}ek, Luca Delacr\'etaz, Yingfei Gu, Duncan Haldane, Bert Halperin, Nick G. Jones, Eslam Khalaf, Max Metlitski, Chaitanya Murthy, Frank Pollmann, Nati Seiberg, Hassan Shapourian, Robert-Jan Slager, Ryan Thorngren, Romain Vasseur, Sagar Vijay and Ashvin Vishwanath for discussions. In particular, I thank Tom\'a\v{s} for making me realize the importance of the smoothness condition,
Ryan for suggesting the link with the parity anomaly,
Eslam for suggesting to look at local DOS,
and Ashvin, Frank and Nick for comments on the manuscript. I am indebted to Nick, Ryan and Frank for collaboration on related works \cite{Verresen18,Jones19,Verresen19}. I am supported by the Harvard Quantum Initiative Postdoctoral Fellowship in Science and Engineering and by a grant from the Simons Foundation (\#376207, Ashvin Vishwanath).

%\bibliography{bib.bib}
\bibliography{arxiv.bbl}

\section*{Methods} \label{sec:methods}

{\small \textbf{One-dimensional lattice models (AIII and BDI class).} Consider the lattice fermions $c_{\lambda,n}$ with index $\lambda=A,B$. If these are complex fermions, we have $\{ c_{\lambda,n}^\dagger, c_{\mu,m}^{\vphantom \dagger}\} = \delta_{\lambda \mu} \delta_{n m}$.
If these are Majorana fermions, they are hermitian and $\{ c_{\lambda,n}, c_{\mu,m}^{\vphantom \dagger}\} = 2\delta_{\lambda \mu} \delta_{n m}$. In both cases, the symmetry we enforce is anti-unitary, squares to unity, and maps $c_{A,n} \mapsto c_{A,n}^\dagger$ and $c_{B,n} \mapsto -c_{B,n}^\dagger$; in the complex fermion case (AIII class), this has interpretation of a sublattice symmetry, whereas in the Majorana setting (BDI class), it corresponds to spinless time-reversal symmetry (to see this, note that $c_{A,n} = a_{n} + a_{n}^\dagger$ and $c_{B,n} = i(a_{n} - a_{n}^\dagger)$, where $a_n$ are the physical superconducting complex fermions, which are invariant under the symmetry transformation). We define the Fourier modes
\begin{equation}
\begin{array}{ll}
c_{A,k} &= \frac{1}{\sqrt{N}} \sum_n e^{-ikn} c_{A,n}, \\
c_{B,k} &= \frac{i}{\sqrt{N}} \sum_n e^{-ikn} c_{B,n}.
\end{array}
\label{eq:latticemomentum}
\end{equation}
Note the unconventional factor of $i$ in the latter, which is convenient when trying to treat AIII and BDI at the same time. The Hamiltonian is
\begin{equation}
H =  \sum_k \left( c_{A,k}^\dagger, c_{B,k}^\dagger \right) \mathcal H(k) \left(
\begin{array}{c}
c_{A,k} \\ c_{B,k}
\end{array} \right). \label{eq:HamMethods}
\end{equation}
Hermiticity of $H$ implies hermiticity of $\mathcal H(k)$, i.e., $\mathcal H = h_0 I + h_x \sigma_x + h_y \sigma_y + h_z \sigma_z$. We work in the setting where the uninteresting $h_0=0$. The above symmetry implies that $\mathcal H(k) = - \sigma_z \mathcal H(k) \sigma_z$, which enforces $h_z = 0$.

Thus far, we have treated the complex fermion and Majorana case on equal footing. However, in the latter case, we have the extra constraint $c_{A,k}^\dagger = c_{A,-k}$ and $c_{B,k}^\dagger = - c_{B,-k}$. This means that it is sufficient to know $\mathcal H(k)$ on half of the Brillouin zone. More concretely, we derive that $\mathcal H(-k) = \mathcal H(k)^*$. In other words, in the Majorana case, we have that $h_x$ ($h_y$) is even (odd) in $k$. Equivalently, $h(k) := h_x(k) + i h_y(k)$ satisfies $h(k)^* = h(-k)$.

As fixed point examples of the topological phases, consider $h(k) = e^{i\alpha k}$, i.e., $\mathcal H_\alpha(k) = \cos(\alpha k) \sigma_x + i \sin(\alpha k) \sigma_y$, which has winding number $\omega = \alpha$. Plugging this in, we obtain the real-space expressions $H_\alpha = - i \sum_n c_{B,n}^\dagger c_{A,n+\alpha}^{\vphantom \dagger} + h.c.$. (In the Majorana case, this can be simplified to $H_\alpha = -2i \sum_n c_{B,n} c_{A,n+\alpha}$.) For $\alpha=1$, we recognize these as being (proportional to) the Su-Schrieffer-Heeger\cite{Su79} and Kitaev\cite{Kitaev01} chain. More generally, we see that there are $|\alpha|$ zero-energy edge modes of $A$-($B$-)type on the left (right) edge for $\alpha>0$ (and oppositely for $\alpha<0$).

The Hamiltonians depicted in Fig.~\hyperref[fig:winding]{2a--e} are (in order of appearance): $\frac{1}{9} H_1 - H_0$, $H_1- H_0$, $H_1$, $H_1 - H_2$ and $\frac{1}{9}H_1 - H_2$. The multi-critical point in Fig.~\hyperref[fig:winding]{2f} is $2H_1 - H_0 -H_2$. For more on these lattice models and their phase diagrams, see Ref.~\onlinecite{Verresen18}.

\textbf{Topological invariant.} We generalize the usual winding number to the case where $\bm h = (h_x,h_y)$ has an arbitrary number of gapless points $\{k_i\}_i$:
\begin{equation}
\omega = \frac{1}{2\pi} \lim_{\delta \to 0} \int_{ \forall i: |k-k_i| > \delta} \partial_k \textrm{arg} (\bm h) \; \mathrm dk. \label{eq:defomegamethods}
\end{equation}
For notational convenience, we consider the case with a single gapless point, which we set at $k_0 = 0$ (the more general case follows straightforwardly). Moreover, let us use the above $h(k) := h_x(k) + i h_y(k)$.
As mentioned in the main text, we presume that $h(k)$ is continuous everywhere and that there is a (non-negative) integer $n$ such that $(\partial_k^n h)(0)$ exists and is finite; let $n$ be the smallest such integer. (The case $n=0$ is the usual gapped case.) By l'H\^{o}pital's rule, $h(k) = c(k) \; k^n$
for $k \in [-\delta,\delta]$, with $c(k)$ continuous and non-vanishing. Define
\begin{equation}
h_\pm(k) := \left\{ \begin{array}{ccl}
h(k) && \textrm{if } |k| \geq \delta \\
c(k) \; \left( k \pm i (k^2-\delta^2) \right)^n &&\textrm{if } |k| < \delta.
\end{array} \right.
\end{equation}
Observe that $h_\pm$ is continuous and gapped/non-vanishing. Hence, its winding number is an integer $m_\pm$. Recalling the definition \eqref{eq:defomegamethods}, we thus have
\begin{align}
m_\pm &= \omega + \frac{1}{2\pi} \lim_{\delta \to 0} \int_{-\delta}^{\delta}\partial_k \arg \left( c(k) \left( k \pm i (k^2-\delta^2) \right)^n \right)  \mathrm dk \notag \\
&= \omega + \frac{n}{2\pi} \int_{-\delta}^{\delta}\partial_k \arg \left( k \pm i (k^2-\delta^2) \right) \mathrm dk = \omega \pm \frac{n}{2}.
\end{align}
Firstly, this tells us that $\omega$ in Eq.~\eqref{eq:defomegamethods} is a (half-)integer. Secondly, $\omega = \frac{m_+ + m_-}{2}$: it is the average of the winding numbers of the two gapped phases created by a particular perturbation. Note that all gapped phases with winding numbers $m_- < m < m_+$ are also proximate to this critical point, which can be included without changing the average (since $\frac{a+(a+1)+\cdots+(b-1)+b}{b-(a-1)} = \frac{a+b}{2}$). Thirdly, $m_+ - m_- = n$. It is not hard to see that this is the maximal difference in winding number obtainable from a gapless point with $h(k) \sim k^n$.

We thus see that $\omega$ is a well-defined, quantized topological invariant, conditional on $(\partial_k^n h)(0)$ being well-defined and nonzero. Fortunately, this is a very natural condition: smoothness in momentum space relates to locality in real space. Indeed, even for usual topological invariants one requires continuity, which corresponds to real-space hoppings decaying faster than $1/r^{d}$ (in $d$ spatial dimensions). Similarly, the aforementioned smoothness condition requires a slightly stricter locality: hoppings decaying faster than $1/r^{d+n}$. The reason for the $n$ dependence is that this is the threshold for locality in the emergent low-energy field theory $H = i \int \chi(x) \partial^n \tilde \chi(x) \mathrm d^d x$. Indeed, the non-local perturbation $\lambda \iint \frac{ \chi(x) \tilde \chi(y)}{(x-y)^a} \mathrm d^d x \mathrm d^d y$ has scaling dimension $[\lambda]=d+n-a$, which is thus irrelevant if $a>d+n$.

\textbf{Edge modes at spatial interfaces.} For the interface from $\omega = 0$ to $\omega=1$, we used the model in Eq.~\eqref{eq:gapped} with $D_x = \partial_x + m(x)$. In Eq.~\eqref{eq:jackiwrebbi}, we found the localized mode $\psi(x) = e^{-\int_0^\infty m(x')\mathrm dx'}$. This can be evaluated exactly for the special case
\begin{equation}
m(x) = a \tanh(x)\qquad \Rightarrow \qquad \psi(x) = \frac{1}{\cosh^a(x)}.
\end{equation}
Figure \hyperref[fig:domain]{3b} plots this for $a=2$. For the interface from $\omega=0$ to $\omega=\frac{3}{2}$, we used the model in Eq.~\eqref{eq:criticalinterface}. In Fig.~\hyperref[fig:domain]{3c}, we show the numerical solution for $\psi(x)$ with $\kappa(x) = 2\tanh(x)$ and $m(x) = 1.05 (\tanh(x)-1)$.

However, we also know on general grounds that a localized zero-energy mode exists for Eq.~\eqref{eq:criticalinterface}. The argument goes as follows: firstly, we know that the second-order differential equation $D_x \psi(x) = 0$ has two solutions. Secondly, on the asymptotic right, we have $D_x = -\partial^2 - \kappa \partial = - \partial(\partial+\kappa)$, which has zero-energy wavefunctions $\psi_1(x) \sim 1$ and $\psi_2(x) \sim \exp(-\kappa x)$. We choose the solution with the latter decay. Thirdly, on the asymptotic left, \emph{both} solutions (where $D_x = -\partial^2 - \kappa \partial + m$ with $\kappa,m<0$) are exponentially decaying (see both signs in Eq.~\eqref{eq:psicrit}). This is essential, since we have chosen the behavior on the right, we have no more further freedom in eliminating solutions.

This argument readily extends to general differential operators $D_x$. We sketch the ingredients here, which we flesh out in more detail in Appendix~\ref{app:indextheorem}.
%the Supplemental Materials\cite{Suppl}.
Let $N$ be the order of $D_x$. Hence, there are $N$ solutions to the differential equation $D_x \psi(x) = 0$.
We now want to exclude the solutions which are \emph{not} exponentially decaying at infinity.
To make this precise, we work in the setting where the coefficients are asymptotically constant, such that we have well-defined characteristic polynomials there.
Roots with negative real part correspond to exponentially decaying solutions on the right.
Hence, the number of roots with \emph{non-negative} real part give us the number of constraints on the right. Similarly on the left. Subtracting these constraints from $N$ gives us a lower bound on the number of localized zero-energy modes! Moreover, this number can only change if we tune one of the roots of the characterisic polynomials to be purely imaginary. But at this point, it corresponds to an additional bulk zero-energy mode, i.e., we would have changed the bulk criticality. See Appendix~\ref{app:indextheorem} for more details.

\textbf{Lattice models and LDOS for Chern transitions.} For each $\alpha \in \mathbb Z$, we define the two-band model
\begin{align}
\mathcal H_\alpha(\bm k) = 
\left\{ \begin{array}{c}
\sin(\alpha k_x) \sigma_x - \sin(k_y) \sigma_y\\
+ (1-\cos(\alpha k_x)-\cos(k_y))\sigma_z.
\end{array} \right.
\end{align}
One can check that for any $\alpha$, this is gapped and the Chern number is $\mathcal C=\alpha$. We consider interpolations
\begin{equation}
\mathcal H (\bm k) = \frac{a \mathcal H_0 + b \mathcal H_1 +c \mathcal H_2}{a+b+c} \qquad \textrm{(with $a,b,c>0$).}
\end{equation}
The complete phase diagram is shown in
%the Supplemental Materials\cite{Suppl}
Appendix~\ref{app:Chernlattice}. For Fig.~\ref{fig:exp}, we focus on the following five points in parameter space:
\begin{equation}
(a,b,c) = \left\{ \begin{array}{lll}
(4,1,1) & & \textrm{has } \mathcal C = 0, \\
(5,6,1) & & \textrm{has } \mathcal C = \frac{1}{2}, \\
(1,5,1) & & \textrm{has } \mathcal C = 1, \\
(1,6,5) & & \textrm{has } \mathcal C = \frac{3}{2}, \\
(1,1,4) & & \textrm{has } \mathcal C = 2.
\end{array} \right.
\end{equation}
For system sizes $L_x \times L_y$ with $L_x = 300$ and $L_y = 1000$ we calculated the local density of states (LDOS) with periodic (open) boundary conditions in the $y$-($x$-)direction. This was done by making a histogram in energy space, where for each energy eigenstate we calculated the density on two strips (along the $y$-axis) in the middle of the system (to obtain the bulk LDOS) and two strips along the left edge (to obtain the edge LDOS).

In
%the Supplemental Materials\cite{Suppl}
Appendix~\ref{app:Chernlattice}, we also use these lattice models to confirm the analytic predictions about stability of the chiral edge mode at the Chern transitions (derived in Appendix~\ref{app:Chernfield}).

\textbf{Edge modes in second quantization.} It is useful to reformulate the approach of the main text in the framework of second quantization. This is indispensable for studying the stability to interactions. In addition, we will find it to be useful to work with a hard boundary rather than the spatial interface used in the main text. Such a hard boundary---and the corresponding boundary condition---can be derived from the spatial interface by taking the RG fixed point limit of the vacuum, as we explain here. These boundary conditions are essential for subsequently studying the effect of interactions.

We illustrate all this in the one-dimensional Majorana case (i.e., BDI class). In particular, this means that we can simplify Eq.~\eqref{eq:HamMethods} as $H = - 2 \sum_k h(k) c_{B,-k} c_{A,k}$, where $h := h_x + i h_y$ as before.
Consider the continuum fields $\chi(x) = \frac{1}{\sqrt{a}} c_{A,n}$ and $\tilde \chi(x) = \frac{1}{\sqrt{a}} c_{B,n}$, where $a$ is the lattice spacing. These hermitian fields obey $\{\chi(x),\chi(y)\} = 2\delta(x-y)$ and $\{\chi(x),\tilde \chi(y)\}=0$. If we define their Fourier modes as $\chi_k = \frac{1}{\sqrt{2\pi}} \int e^{-ikx}\chi(x)\mathrm dx$ and $\tilde \chi_k = \frac{1}{\sqrt{2\pi}} \int e^{-ikx}\tilde \chi(x)\mathrm dx$, then these relate to the modes defined in Eq.~\eqref{eq:latticemomentum} as $c_{A,a k} = \sqrt{\frac{2\pi}{aN}} \chi_k$ and $c_{B,a k} = i \sqrt{\frac{2\pi}{aN}} \tilde \chi_k$. Using the correspondence $\frac{1}{N} \sum_k \leftrightarrow a \int \frac{\mathrm d k}{2\pi}$, we have
\begin{equation}
H \propto i \int_{-\infty}^{+\infty} \tilde \chi_{-k} h(k) \chi_k \mathrm d k = i \int_{-\infty}^{+\infty} \tilde \chi(x) D_x \chi(x) \mathrm d x, \label{eq:continuumH}
\end{equation}
where $D_x = h(-i\partial_x)$. The unspecified prefactor is $-2$, which we will suppress for convenience (note that in bipartite systems one can absorb overall signs). We now discuss the emergence of boundaries.

\textbf{Boundary conditions.} To study edge modes and their stability, it is crucial to also know the boundary conditions of the above field theories. There are different ways of obtaining this\cite{Cho17,Oshikawa19}, but here we follow an intuitive domain-wall approach. (See Appendix~\ref{app:bc} for more details.) In the main text, we saw that for the critical point with $\omega=\frac{1}{2}$, we had $D_x = \partial_x +m$, i.e., $H = i \int \tilde \chi(\partial + m) \chi \mathrm dx$ with $m < 0$ being trivial. We can interpret a half-infinite chain ($x>0$) with a hard boundary at $x=0$ as arising from an infinitely-long chain with a mass profile $m(x)$ where $m(x) \to - \infty$ for $x<0$. Before taking this limit, let us define $\chi_\textrm{loc} = \int_{-\infty}^{0} e^{-\int_0^x m(x') \mathrm d x'} \chi(x) \mathrm d x$. We claim that $[\chi_\textrm{loc},H] = 2i\tilde \chi(0)$. If we now take the limit of $m(x) \to -\infty$ (for $x<0$), then $\chi_\textrm{loc} \to 0$. Hence, its commutator must vanish, deriving the boundary condition $\tilde \chi(0) = 0$. (Similarly, one can derive that $\chi(L)=0$ at the other edge.)

To derive that $[\chi_\textrm{loc},H] = 2i\tilde \chi(0)$, it is useful to first partially integrate such that $H = \int_{-\infty}^{\infty} \chi (\partial - m) \tilde \chi \mathrm d x$ and to note that
\begin{equation}
[\chi(x),\chi(y) \tilde D \tilde \chi(y)] = \{\chi(x),\chi(y)\} \tilde D_y \tilde \chi(y) = 2\delta(x-y)\tilde D \tilde \chi(y).
\end{equation}
Altogether, we thus have that
\begin{equation}
\left[ \chi_\textrm{loc},H \right] = 2i \int_{-\infty}^0 e^{-\int_0^x m(x')\mathrm dx'} (\partial-m) \tilde \chi(x) \mathrm d x.
\end{equation}
The claimed result then follows from partial integration.

Some methods only derive that, e.g., the product $\chi(0) \tilde \chi(0)$ has to vanish at the left edge---which guarantees that integration by parts does not generate boundary terms. However, for us it is very important to know which of the two fields actually vanishes (as we will see when discussing the stability of edge modes). More conceptually, the Hamiltonian $H = i \int_0^\infty \tilde \chi (\partial + m) \chi \mathrm d x$ only tells us that $m>0$ has one higher winding number than $m<0$. The boundary condition $\tilde \chi(0)=0$ fixes the winding number of $m<0$ to be $\omega = 0$ (whereas the alternative condition $\chi(0) = 0$ would mean that $\omega = -1$ for $m<0$).

The above illustrated the derivation of boundary conditions for a particular choice of $D_x$. In
%the Supplemental Materials
Appendix~\ref{app:bc}, we prove a more general theorem for this purpose. %\cite{Suppl}.
Using this, we derive that the model in Eq.~\eqref{eq:criticalinterface} has boundary condition $\tilde \chi(0) = \partial_x \tilde \chi(0) = 0$.

\textbf{Edge modes for half-infinite systems.} Having derived the boundary condition, we can now study the edge modes in second quantization. For instance, for the phase diagram in Fig.~\hyperref[fig:domain]{3a}, we have the half-infinite system
\begin{equation}
H = i \int_{0}^{+\infty} \tilde \chi \left( - \partial^2 - \kappa \partial + m \right) \chi \mathrm dx \textrm{ with }\tilde \chi(0) = \partial_x \tilde \chi(0) = 0. \label{eq:Methodscrit}
\end{equation}
Let us now study the critical system with $\omega=\frac{3}{2}$, i.e., $m = 0$ and $\kappa >0$. This has a localized zero-energy mode $\chi_\textrm{loc} = \int_0^\infty e^{-\kappa x} \chi(x) \mathrm d x$. Indeed, one straightforwardly derives that
\begin{equation}
[\chi_\textrm{loc},H] = 2i \int_0^\infty e^{-\kappa x} \left( -\partial + \kappa  \right) \partial \tilde \chi \mathrm d x = 2 i \partial \tilde \chi(0) = 0,
\end{equation}
which vanishes by the above boundary condition.

So far we have only taken the RG limit of the trivial vacuum, i.e., the `outside'. It is convenient to also take the RG limit of our critical system. In this case, since $\kappa \neq 0$, the quadratic term in Eq.~\eqref{eq:Methodscrit} is RG-irrelevant. We thus flow to
\begin{equation}
H = i \int_0^\infty \tilde \chi(x) \partial_x \chi(x) \mathrm d x \quad \textrm{with} \quad \tilde \chi(0) = 0, \label{eq:MethodsRG}
\end{equation}
where we have suppressed the overall constant $-\kappa$. However, the price we pay for working in this fixed point limit, is that we need to treat $\chi_\textrm{loc}$ as a separate mode (which has pinched off into an exactly-localized, zero-energy degree of freedom in this RG limit). Without this mode, Eq.~\eqref{eq:MethodsRG} is just the usual $\omega=\frac{1}{2}$ transition. Indeed, one can think of the $\omega=\frac{3}{2}$ case as $\omega=\frac{1}{2}$ plus a localized mode. This is also clearly illustrated in the lattice models whose hoppings are shown in Fig.~\ref{fig:winding}.

We have thus reduced the study of $\omega=\frac{3}{2}$ to a study of Eq.~\eqref{eq:MethodsRG} and a decoupled edge mode $\chi_\textrm{loc}$. From this perspective, the non-trivial part is to argue that this edge mode remains localized after coupling it to the critical bulk. The dominant couplings would be $V = i \chi_\textrm{loc} \chi(0)$ or $V = i  \chi_\textrm{loc} \tilde \chi(0)$. However, the former is not allowed by symmetry, whereas the latter is zero by virtue of the boundary condition!

Hence, the dominant coupling is $V = i \chi_\textrm{loc} \partial_x \tilde \chi(0)$. Using the scaling dimensions $[\chi_\textrm{loc}] = 0$ and $[\tilde \chi] = \frac{1}{2}$, we see that this coupling $\lambda$ in $H+\lambda V$ has dimension $[\lambda] = -\frac{1}{2}$ (since $[\lambda V]=1$ has units of energy). In other words, this coupling is RG-irrelevant, implying that the edge mode survives. However, this is not the full story: it turns out the edge mode remains exactly-localized under this coupling! In particular,
\begin{equation}
\chi_\textrm{loc}' := \chi_\textrm{loc} - \lambda \chi(0) \quad \Rightarrow \quad [\chi_\textrm{loc}',H +\lambda V] = 0,
\end{equation}
which can be verified by a direct computation. The general mechanism or reason behind this is that---despite the coupling $V = i \chi_\textrm{loc} \partial_x \tilde \chi(0)$ being allowed---it can in fact be rotated away by a unitary. More precisely,
\begin{equation}
U(\alpha) :=e^{\frac{\alpha }{2}\chi_\textrm{loc} \chi(0)} \quad \Rightarrow \quad U(\lambda)^\dagger H U(\lambda) = H + \lambda V + O(\lambda^2), \label{eq:rotation}
\end{equation}
using that $\left[ \chi_\textrm{loc} \chi(0) , H\right] = \chi_\textrm{loc} \left[ \chi(0), H \right] = -2i\chi_\textrm{loc} \partial \tilde \chi(0)$. In other words, Eq.~\eqref{eq:rotation} tells us that the effective perturbation secretly has a higher scaling dimension. I.e., effectively we are only perturbing with $V= i \chi_\textrm{loc} \partial_x^2 \tilde \chi(0)$, but this can similarly be rotated away by using $\chi_\textrm{loc} \partial_x \chi(0)$ as a generator. This continues ad infinitum, recovering the fact that free-fermion perturbations are not able to give the edge mode an algebraic dressing!

\textbf{Effect of interactions.} The above formulation perfectly lends itself to study the effect of interactions. The goal is to find the dominant term coupling a localized edge mode $\chi_\textrm{loc}$ to the critical bulk in Eq.~\eqref{eq:MethodsRG}. In the non-interacting case (i.e., quadratic couplings) we found that we could rotate away all perturbations, implying that there are no algebraic contributions.

Intriguingly, the mechanism of rotating away perturbations also applies to interacting (say, quartic) couplings. This has been studied in detail in Ref.~\onlinecite{Verresen19}, where it was found that the dominant interacting coupling which could no longer be rotated away is a quartic terms with six derivatives. I.e., the operator has dimension $0 + 3\times \frac{1}{2}+6\times 1 = 7.5$. Hence, the coupling constant $\lambda$ in the perturbed Hamiltonian $H + \lambda V$ has dimension $[\lambda] = -6.5$, being very RG-irrelevant. Since the finite-size energy splitting is quadratic in $\lambda$, i.e, $\sim \lambda^2/L^\beta$ (due to both edges needing to couple to the critical bulk), we find $\beta = 2[\lambda]-1 = 14$. This same exercise could be repeated for any critical point of interest.

%\textbf{Data and code availability.} The data that support the findings of this study are available from the corresponding author upon request.
}

%\pagebreak

\widetext
\ifx\targetformat\undefined
\begin{center}
	\textbf{\large Supplemental Materials}
\end{center}

\setcounter{equation}{0}
\setcounter{figure}{0}
\setcounter{table}{0}
\makeatletter
\renewcommand{\theequation}{S\arabic{equation}}
\renewcommand{\thefigure}{S\arabic{figure}}
\renewcommand{\bibnumfmt}[1]{[S#1]}
\renewcommand{\citenumfont}[1]{S#1}

\else
\appendix
\fi

\newpage

\section{A generalized index theorem \label{app:indextheorem}}

Consider the differential operator
\begin{equation}
D_x = \sum_{n=0}^N c_n(x) \; \left(\frac{\partial}{\partial x} \right)^n \qquad \textrm{where } c_n(x) \in \mathbb C. \label{eq:Dx}
\end{equation}
We work in the setting where $c_n(x)$ is asymptotically constant for $|x| \to \infty$, denoting these limits as $c_n^\pm := \lim_{x \to \pm \infty} c_n(x)$ (we presume that $c_n(x) - c_n^\pm$ decays to zero quickly enough; see the proof for quantitative statements). Moreover, we presume that the highest-order coefficient $c_N(x)$ never vanishes. One could relax this condition, but one would have to treat such vanishing points as defects with particular boundary conditions, complicating the analysis.

We define the asymptotic symbols or characteristic polynomials $f_\pm(z) = \sum_{n=0}^N c_n^\pm z^n$.
Note that these contain all the physical information of the asymptotic regions. In particular, the energy spectrum is given by $|f_\pm(ik)|$, telling us that the region $x \to \pm \infty$ gapped if and only if $f_\pm$ has no purely-imaginary roots.

\begin{shaded}
\begin{theorem*}
Let $N_\pm$ denote the number of roots of $f_\pm(z)$ with strictly negative real part. If the region $x \to -\infty$ is gapped (i.e., $f_-(z)$ has no purely-imaginary roots), then
\begin{equation}
\# \left( \textrm{ exponentially-localized solutions of } D_x \psi(x) = 0 \; \right) \qquad \geq \qquad N_+ - N_-. \label{eq:theorem} 
\end{equation}

If $N_+ - N_- > 0$, the wavefunctions have the asymptotic form $\psi_i(x) \sim \exp\left(z_i x\right)$ ($i=1,\cdots, N_+ - N_-$) for $x\to + \infty$, where $\{z_i\}_{i}$ are the $N_+ - N_-$ roots of $f_+(z)$ with negative real part closest to the imaginary axis. %with negative $\textrm{Re } z_i$.

Moreover, $N_+ - N_-$ cannot change without changing the universality class of the asymptotic regions $x\to \pm \infty$.
\end{theorem*}
\end{shaded}

\begin{figure}%[h]
	\includegraphics[scale=1]{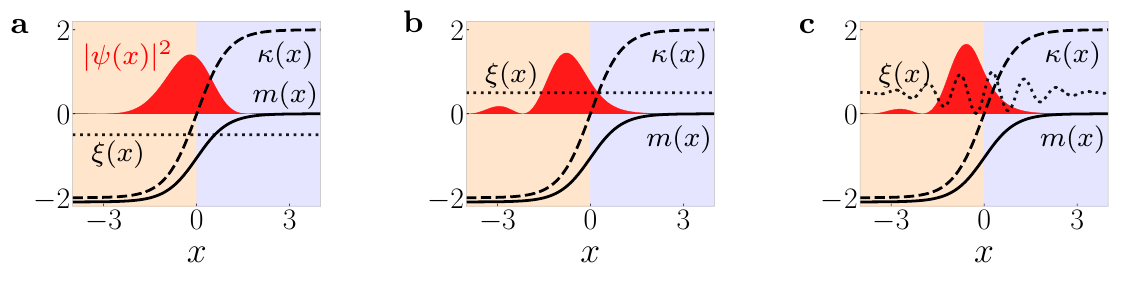}
	\caption{Interfaces between $\omega = 0$ and $\omega=\frac{3}{2}$, now including a third-order derivative with prefactor $\xi(x)$. In all three cases, we take $\kappa(x) = 2 \tanh(x)$ and $m(x) = 1.05(\tanh(x)-1)$ and numerically solve for the zero-energy state $\psi(x)$. \textbf{a}, $\xi(x) = -0.5$. \textbf{b}, $\xi(x) = 0.5$. \textbf{c}, $\xi(x) = 0.5 \left( 1 + \sin(6x) \exp\left(-x^2/4\right) \right)$.  \label{fig:stability}}
\end{figure}

This theorem can be seen as a continuum analogue of Theorem 1 in Ref.~\onlinecite{Verresen18}. Before proving it, we apply this theorem to some examples:
\begin{enumerate}
	\item $D_x = \partial_x + m(x)$ where $m^\pm = \lim_{x \to \pm \infty} m(x) = \pm |m_\infty|$. We thus have $f_\pm(z) = z \pm |m_\infty|$. This implies that $N_+ = 1$ and $N_- = 0$, i.e., we have one edge mode, which decays as $\exp(-|m_\infty|x)$ for $x\to +\infty$. (In this particular case, we have the closed form\cite{Jackiw76} $\psi(x) = \exp\left( - \int_0^x m(x') \mathrm dx' \right)$.)
	\item $D_x = -\partial_x^2 - \kappa(x) \partial + m(x)$ with $\kappa^\pm=\lim_{x \to \pm \infty} \kappa(x) = \pm |\kappa_\infty| \neq 0$ and $\lim_{x\to +\infty} m(x) = 0$ and $\lim_{x \to - \infty} m(x) = - |m_\infty|$. We thus have $f_+(z) = -z^2 - |\kappa_\infty|z$, with roots $z_1 = 0$ and $z_2 = -|\kappa_\infty|$, i.e., $N_+ = 1$. Similarly, one derives that $N_- = 0$, so we again have one edge mode, now decaying as $\exp(-|\kappa_\infty|x)$ for $x\to +\infty$.
	\item $D_x = \xi \partial_x^3 -\partial_x^2 - \kappa(x) \partial + m(x)$ where $\kappa(x)$ and $m(x)$ have the same conditions as in the previous examples. Note that the limit $\xi \to 0$ recovers the previous example; indeed, the purpose of this example is to show the edge mode's stability to turning on a higher-order derivative. We have $f_+(z) = (\xi z^2 -z - |\kappa_\infty|)z$. For $\xi \approx 0$, this has roots $z_1 = 0$, $z_2 \approx - |\kappa_\infty|$ and $z_3 \approx 1/\xi$. Hence, if $\textrm{Re } \xi > 0$, we have $N_+ = 1$, whereas for $\textrm{Re } \xi < 0$ we have $N_+ = 2$. Similarly, one derives that $\textrm{Re } \xi > 0$ ($\textrm{Re } \xi < 0$) implies $N_- = 0$ ($N_- = 1$) if $\xi$ is small. Hence, for small $\xi$, we find a localized edge mode which decays into the critical bulk with $\psi(x) \sim \exp(-|\kappa_\infty|x)$ for $x\to +\infty$. (Note that $\xi$ has to have a finite real value, otherwise we have an additional critical mode in the bulk; in particular, it would not allow us to connect to the limit where $\xi = 0$.) Examples are shown in Fig.~\ref{fig:stability}.
	
	This example also nicely illustrates two important subtleties:
	\begin{enumerate}
		\item What about larger $\xi$? In particular, one can check that for large positive $\xi$, we have $N_- = 2$, such that the lower bound $N^+ - N^- = -1$ no longer ensures the existence of an edge mode! However, this region is separated from small $\xi$ by a phase transition in the asymptotic region $x \to - \infty$. For instance, for $\xi = |\kappa_\infty| = |m_\infty| = 1$, we have $D_x = \partial^3 - \partial^2 + \partial - 1 = (\partial-1)(\partial^2 + 1)$. Its spectrum (given by replacing $\partial \to ik$) is $|(ik-1)(1-k^2)|$, which is gapless for $k = \pm 1$. I.e., this shows that one way of destroying the edge mode is by driving the trivial vacuum through a quantum phase transition, which is of course not surprising.
		\item What about a spatially-dependent $\xi(x)$? In particular, the above argument seemed to rely on the signs of $\lim_{x \to \pm \infty} \textrm{Re } \xi(x)$ agreeing. Indeed, if the signs were different, the edge mode could disappear. However, this can again not be connected to the case above. If $\xi(x) \in \mathbb R$, then the asymptotic signs must agree since we do not allow $\xi(x)$ to vanish for any $x$ (see the comments below Eq.~\eqref{eq:Dx}).
		If $\xi(x) \in \mathbb C$, one can indeed have opposite signs for $\lim_{x \to \pm \infty} \textrm{Re }\xi(x)$, but it is separated by the aforementioned case by a phase transition where $\xi$ becomes purely imaginary. Relatedly, in cases where $\xi(x)$ is purely-imaginary for some value of $x$, we cannot consider the limit $\xi \to 0$ since for that value of $x$, this perturbation is not RG-irrelevant.
	\end{enumerate}
\end{enumerate}

\begin{proof}
Due to the fundamental theorem of algebra, we can write
\begin{equation}
f_\pm(z) = \alpha_\pm \prod_{m=1}^{M_\pm} \left( z-z_{m}^\pm \right)^{\mu_{m}^\pm}
\end{equation}
where $M_\pm$ counts the number of distinct roots and $\mu_n^\pm$ their multiplicities, i.e., $\sum_{m=1}^{M_\pm} \mu_m^\pm = N$. We expect that these roots dominate the asymptotic behavior of the solutions to $D_x \psi(x) = 0$ for large $|x|$. This can be made precise: for either $x \to + \infty$ or $x \to - \infty$, there is a basis of solutions which have the asymptotic form
\begin{equation}
\psi^{\pm}_{m,i}(x) = x^i \; \exp\left( z^\pm_m x \right) \; \left( 1 + o(1) \right) \textrm{ as } x \to \pm \infty, \qquad \textrm{for } m =1,\cdots, M_\pm \textrm{ and } i = 1, \cdots, \mu_m^\pm. \label{eq:solutions}
\end{equation}
This follows from the theorems in Refs.~\onlinecite{Levinson48,Devinatz65} under the assumption that $c_n(x) = c_n^\pm + O(1/x^{q_\pm})$ as $x\to \pm \infty$, with $q_\pm> 2\; \textrm{max}_m \mu_m^\pm -1$. (It is likely that $q>1$ is sufficient if all one cares---as we do here---about the exponential factor.)

The sets of states $\{ \psi^+_{m,i} \}_{m,i}$ and $\{ \psi^-_{m,i} \}_{m,i}$ both independently form a basis for the $N$-dimensional space of solutions $V$. The former (latter) tells us how the solutions behave asymptotically to the right (left). Since we are interested in solutions which decay exponentially, let $V_+ \subset V$ ($V_- \subset V$) be the subspace of states which decay exponentially to the right (left). From Eq.~\eqref{eq:solutions}, we see that $V_+$ is generated by the states with $\textrm{Re } z_m^+ <0$, and $V_-$ by those with $\textrm{Re } z_m^- >0$, respectively. Hence, $\dim V_+ = N_+$ and $\dim V_- = N - N_-$ (using the assumption that all roots of $f_-$ have non-zero real part). The intersection of these two spaces contain solutions which decay exponentially to both sides: these are our desired localized zero-energy solutions. Note that
\begin{equation}
\dim \left( V_+ \cap V_- \right) = \dim V_+ + \dim V_- - \dim\left( V_+ + V_- \right) \geq N_+ + (N-N_-) - N = N_+ - N_-
\end{equation}
which proves the claimed inequality Eq.~\eqref{eq:theorem}.

Moreover, we have also proven (by construction) the claim about their asymptotic form.

Lastly, it is clear that $N_+ - N_-$ can only change by either changing $N_+$ or $N_-$, and by continuity of roots, this would require a root becoming purely-imaginary. However, this would correspond to a bulk gapless mode, changing the universality class of the asymptotic regions. (Note that even tuning on a higher-order derivative $\varepsilon \partial^{N+1}$ in $D_x$ cannot change $N_+ - N_-$ since it would change $N_+$ and $N_-$ by equal amounts.)
\end{proof}

\section{Edge modes at critical Chern insulators: field theory \label{app:Chernfield}}

Here, we study the stability of a chiral edge mode at a Chern insulator phase transition. Our set-up is to take the phase transition between Chern numbers $\mathcal C = 0\leftrightarrow \mathcal C=1$ and to bring in an additional initially-decoupled chiral edge mode (taken together, this is thus in the same class as the transition between $\mathcal C = 1\leftrightarrow \mathcal C=2$). Moreover, we work in second-quantized notation. See the \hyperref[sec:methods]{Methods} for general comments on how such a set-up is equivalent to the interface set-up studied in the main text.

The bulk critical (Dirac cone) Hamiltonian which we define can be interpreted as a coupled-wire construction of one-dimensional chiral modes (with the fields $\psi$ and $\varphi$ denoting the two chiralities):
\begin{align}
H_0 &= \iint \left( \psi^\dagger (-i\partial_\para) \psi +  \varphi^\dagger (i\partial_\para) \varphi - \psi^\dagger \partial_\perp \varphi + \varphi^\dagger \partial_\perp \psi \right) \mathrm dx_{\para}\mathrm d x_{\perp} \\
&= \iint \left( \psi^\dagger, \varphi^\dagger \right)
\left( \sigma^z (- i \partial_\para) + \sigma^y (-i \partial_\perp) \right)
\left( \begin{array}{c}
\psi\\
\varphi
\end{array} \right) \mathrm dx_{\para} \mathrm d x_{\perp} \label{eq:sigma} \\
&=  \iint \left( k_\para \psi^\dagger \psi - k_\para \varphi^\dagger \varphi - \psi^\dagger \partial_\perp \varphi + \varphi^\dagger \partial_\perp \psi \right)\mathrm dk_{\para} \mathrm d x_{\perp} .
\end{align}
With periodic boundary condition, the (two-band) spectrum is $\varepsilon = \pm \sqrt{ k_\para^2 + k_\perp^2}$, confirming the claim that this is a Dirac cone.

Throughout we will use the notation $x_\para$ and $x_\perp$ to denote the coordinates parallel and perpendicular to the boundary at $x_\perp = 0$. Moreover, it will be useful to work in the above mixed representation of $k_\para$ and $x_\perp$. The boundary condition (at the left edge) is $\varphi|_{x_\perp =0} = 0$ for the transition between the trivial phase and the Chern insulator where the left edge has a mode of the ``$\psi$'' type.

\subsection{Bulk gapped phases and edge modes}
The above boundary condition can be derived from an interface set-up as described in the \hyperref[sec:methods]{Methods}. However, here we simply illustrate that the above choice is indeed the one that guarantees edge modes in the nearby gapped Chern insulator. Moreover, it will be a useful warm-up to derive the edge modes in the gapped case, as the type of manipulations we perform will occur repeatedly in the critical case.

To gap out the system, we need to introduce the $\sigma^x$ component in Eq.~\eqref{eq:sigma}. Hence, the bulk mass term is $H_M = M\iint (\psi^\dagger \varphi + \varphi^\dagger \psi)$. To see the existence of edge modes for $M>0$, define
\begin{equation}
\psi_\textrm{loc}(k_\para) := \int_0^\infty e^{-M x_\perp} \psi(k_\para,x_\perp) \mathrm d x_\perp.
\end{equation}
This is exponentially localized on the left edge, and has a $1+1d$ gapless spectrum:
\begin{align}
\boxed{ [\psi_\textrm{loc}(k_\para),H_0 + H_M] } &=  \int_0^\infty e^{-M x_\perp} \underbrace{\left\{\psi(k_\para,x_\perp),\psi(q_\para,y_\perp)^\dagger \right\}}_{=\delta(k_\para-q_\para)\delta(x_\perp-y_\perp)} \left( k_\para \psi - \partial_\perp \varphi + M \varphi \right) \mathrm d x_\perp \mathrm d q_\para \mathrm d y_\perp \\
&= k_\para \psi_\textrm{loc}(k_\para) + \int_0^\infty e^{-M x_\perp} (M-\partial_\perp) \varphi(k_\para,x_\perp) \; \mathrm d x_\perp \\
&= \boxed{ k_\para \psi_\textrm{loc}(k_\para) }+ \int_0^\infty \underbrace{\left[ (M+\partial_\perp) e^{-M x_\perp} \right]}_{=0} \varphi(k_\para,x_\perp) \; \mathrm d x_\perp + \underbrace{\varphi(k_\para,0)}_{=0}.
\end{align}
We see that it was crucial to use the boundary condition that $\varphi$ vanishes at the left edge.

In what follows, we focus on the bulk critical system ($M=0$).

\subsection{Coupling edge mode to critical system}
Let us now introduce an initially-decoupled chiral mode $\eta(x_\para)$ living at the left boundary, described by the one-dimensional Hamiltonian $\int k_\para \eta(k_\para)^\dagger \eta(k_\para) \mathrm dk_\para$. We see that its dispersion implies that it is of the  ``$\psi$'' type, i.e., it has the same chirality as the edge mode of the nearby gapped Chern insulator. This is very important: it means that the new transition is effectively between $\mathcal C=1 \leftrightarrow \mathcal C=2$ (if we had chosen the opposite sign for the dispersion of $\eta$, we would have instead obtained a transition between $\mathcal C=-1 \leftrightarrow \mathcal C=0$, which does \emph{not} have any stable edge modes). We now show that this edge mode remains localized when coupling it to the critical bulk.

\subsubsection{Coupling to opposite chirality}
We first consider the case where we try to couple the chiral edge mode $\eta(x_\para)$ to the $\varphi$ field. This coupling must necessarily involve $\partial_\perp \varphi$ due to the boundary condition which sets $\varphi(x_\perp = 0) = 0$:
\begin{equation}
V = \int \left( k_\para \eta(k_\para)^\dagger \eta(k_\para) + \lambda \; \eta(k_\para)^\dagger \partial_\perp \varphi(k_\para,0) + h.c.\right) \mathrm d k_\para.
\end{equation}
Note that $\eta$ is only defined in $1+1$ dimensions (e.g., $\eta$ has dimension $1/2$, whereas $\psi$ and $\varphi$ have dimension $1$ when the fields are expressed in real space). From this, we derive the scaling dimension $[\lambda] = -1/2 < 0$, which is RG-irrelevant. However, a stronger statement holds: the edge mode remains exactly localized. To see this, first observe:
\begin{align}
[\eta(k_\para),H_0 + V] &= k_\para \eta(k_\para) + \lambda\; \partial_\perp \varphi(k_\para,0) \\
[\psi(k_\para,x_\perp),H_0 + V] &= k_\para \psi - \partial_\perp \varphi .
\end{align}
Hence, if we define the local mode $\boxed{\psi_\textrm{loc}(k_\para) := \eta(k_\para) + \lambda \psi(k_\para,0)}$, then $\boxed{[\psi_\textrm{loc}(k_\para) , H_0 + V] = k_\para \psi_\textrm{loc}(k_\para)}$.

\begin{figure}
\includegraphics[scale=1]{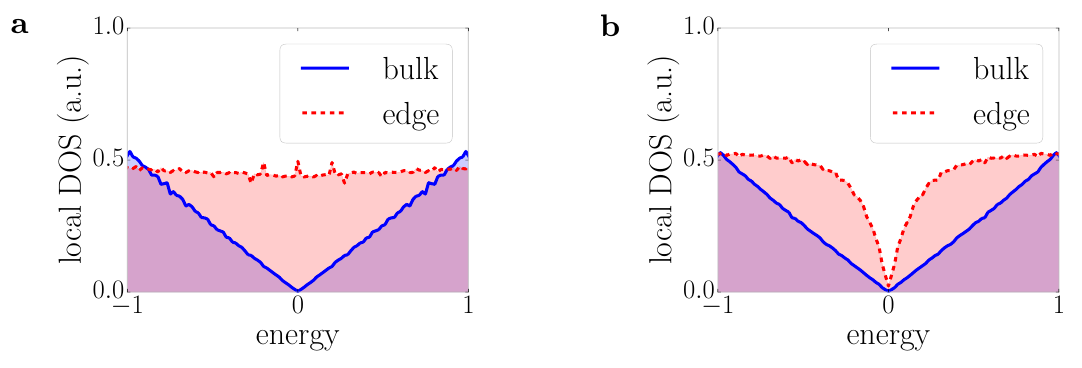}
\caption{Numerically-obtained density of states (DOS) for a discretized field theory where one couples an additional chiral edge mode to a Dirac cone. More particular, we couple to a bulk mode of the \emph{opposite chirality} (for \emph{same chirality}, see Fig.~\ref{fig:samechirality}). \textbf{a}, If we use the boundary condition for a Dirac cone between $\mathcal C=0 \leftrightarrow \mathcal C=1$ (i.e., $\varphi$ vanishes at the left boundary; see the text), the additional edge mode (putting us at a $\mathcal C=1 \leftrightarrow \mathcal C=2$ transition) remains stable as evidenced by the finite DOS. \textbf{b}, To illustrate the importance of using the correct boundary condition, we show how the edge mode disappears if $\psi$ (instead of $\varphi$) is forced to vanish at the boundary. This would correspond to the bulk Dirac cone being a transition between $\mathcal C=-1 \leftrightarrow \mathcal C=0$, such that the additional edge mode puts us at a $\mathcal C=0 \leftrightarrow \mathcal C=1$ transition, which indeed should not have a localized edge mode. (Of course, if we would also swap the chirality of the additional edge mode, we would be at a $\mathcal C=-2 \leftrightarrow \mathcal C=-1$ transition, which again has a stable mode.) \label{fig:latticesamechirality}}
\end{figure}

What if the velocity of the original $\eta$ mode was not the same as the bulk modes? Suppose we instead have $v_\textrm{rel} k_\para \eta^\dagger \eta$, with $v >0$ labeling the (relative) velocity. This does not affect the solution at $k_\para=0$. However, for $k_\para \neq 0$ and $v_\textrm{rel} \neq 1$, obtaining the solution analytically is more challenging. Instead, we numerically confirm the stability. Consider the following lattice discretization of $H_0$ (for a fixed value of $k_\para$):
\begin{equation}
H_0^\textrm{lat}(k_\para) = \sum_{n \geq 1} \left( k_\para \psi^\dagger_n \psi_n^{\vphantom \dagger} - k_\para \varphi^\dagger_n \varphi_n^{\vphantom \dagger} + \left[ \varphi_n^\dagger \left( \psi_{n+1}^{\vphantom \dagger} - \psi_n^{\vphantom \dagger} \right) + h.c.\right] \right).
\end{equation}
For periodic boundary conditions, its dispersion is $\varepsilon_k = \sqrt{k_\para^2 + \sin^2(k_\perp) + (1-\cos(k_\perp))^2} \approx \sqrt{k_\para^2 + k_\perp^2}$ for small $k_\perp \approx 0$. Note that it is crucial that for open boundary conditions, the first site is $\psi_1$ and the second site $\varphi_1$ (this is encoded in the form of the hopping). This ensures that the continuum limit of this model is the above field theory with boundary condition $\varphi(x_\perp = 0) = 0$. As for the perturbation:
\begin{equation}
V^\textrm{lat}(k_\para) = v_\textrm{rel} k_\para \; \eta^\dagger \eta + \lambda \; \eta^\dagger \varphi_1^{\vphantom \dagger} + \bar \lambda \; \varphi_1^\dagger \eta. \label{eq:perturbationsamechiral}
\end{equation}
We now numerically solve for the local density of states, shown in Fig.~\ref{fig:latticesamechirality}(a) for the values $v_\textrm{rel}=2$ and $\lambda=0.5$. This showcases its stability. For comparison, we can consider the lattice model corresponding to $\psi(x_\perp = 0)= 0$:
\begin{equation}
H_0^\textrm{lat}(k_\para) = \sum_{n \geq 1} \left( k_\para \psi^\dagger_n \psi_n^{\vphantom \dagger} - k_\para \varphi^\dagger_n \varphi_n^{\vphantom \dagger} - \left[ \psi_n^\dagger \left( \varphi_{n+1}^{\vphantom \dagger} - \varphi_n^{\vphantom \dagger} \right) + h.c.\right] \right),
\end{equation}
where the leftmost (bulk) site is now $\varphi_1$. In this case, the same perturbation in Eq.~\eqref{eq:perturbationsamechiral} would now correspond to $\eta^\dagger \varphi$ in the continuum (as opposed to $\eta^\dagger \partial_\perp \varphi$). This indeed delocalizes the edge mode as shown in Fig.~\ref{fig:latticesamechirality}(b) (where we choose the same values $v_\textrm{rel}=2$ and $\lambda=0.5$ as before).

\subsubsection{Coupling to same chirality \label{sec:field}}

We now couple the edge mode to a bulk field of the same chirality:
\begin{equation}
V = \int \left( k_\para \eta(k_\para)^\dagger \eta(k_\para) + \lambda \; \eta(k_\para)^\dagger \psi(k_\para,0) + h.c.\right) \mathrm d k_\para. \label{eq:samechirality}
\end{equation}
This is now a relevant perturbation: $[\lambda] = 1/2$. Hence, something drastic will happen. Remarkably, however, we will show that the edge mode persists.

We have
\begin{align}
[\eta(k_\para),H_0 + V] &= k_\para \eta(k_\para) + \lambda\; \psi(k_\para,0) \label{eq:temp1} \\
[\psi(k_\para,x_\perp),H_0 + V] &= k_\para \psi - \partial_\perp \varphi + \bar \lambda \eta \delta(x_\perp) \\
[\phi(k_\para,x_\perp),H_0 + V] &= -k_\para \phi + \partial_\perp \psi .
\end{align}
Let us define the ansatz modes
\begin{equation}
\tilde \psi(k_\para) := \int_0^\infty e^{-a x_\perp} \psi(k_\para,x_\perp)\mathrm d x_\perp \qquad \textrm{and} \qquad
\tilde \varphi(k_\para) := \int_0^\infty e^{-a x_\perp} \varphi(k_\para,x_\perp)\mathrm d x_\perp
\end{equation}
with $a>0$ (such that the modes are localized at the boundary). Then
\begin{align}
[\tilde \psi, H_0 + V] &= k_\para \tilde \psi + \int_0^\infty e^{-ax_\perp} \left( \delta(x_\perp) \bar \lambda \eta - \partial_\perp \varphi \right) \mathrm d x_\perp = k_\para \tilde \psi - a \tilde \varphi + \bar \lambda \eta \\
[\tilde \varphi, H_0 + V] &= -k_\para \tilde \varphi + \int_0^\infty e^{-ax_\perp} \partial_\perp \psi \mathrm d x_\perp = -k_\para \tilde \varphi + a \tilde \psi - \psi(k_\para,0). \label{eq:temp2}
\end{align}
We can eliminate $\psi(k_\para,0)$ by combining Eqs.~\eqref{eq:temp1} and \eqref{eq:temp2} to get
\begin{align}
[\eta + \lambda \tilde \varphi, H_0 + V] &= k_\para \eta - k_\para \lambda \tilde \varphi + a \lambda \tilde \psi \\
[\tilde \psi, H_0 + V] &=  k_\para \tilde \psi - a \tilde \varphi + \bar \lambda \eta.
\end{align}
We now want to find a choice of $\alpha$ such that $\eta + \lambda \tilde \varphi + \alpha \lambda \tilde \psi$ is closed under the commutator:
\begin{equation}
[\eta + \lambda \tilde \varphi + \alpha \lambda \tilde \psi, H_0 + V] = (k_\para + \alpha |\lambda|^2) \eta - (k_\para + \alpha a ) \lambda \tilde \varphi + \left(k_\para + \frac{a}{\alpha}\right) \alpha \lambda \tilde \psi. \label{eq:condition}
\end{equation}
Hence, we get a localized mode with energy $\varepsilon_{k_\para} = k_\para + \alpha |\lambda|^2$ (implying that $\alpha$ is real) if the three coefficients all agree on the RHS. I.e., we obtain the two conditions
\begin{equation}
2k_\para + \alpha |\lambda|^2 + \alpha a = 0 \qquad \textrm{and} \qquad \alpha^2 |\lambda|^2 = a. \label{eq:twoconditions}
\end{equation}
Note that there is no solution for $k_\para = 0$! It would imply that $a = -|\lambda|^2$ or $a=0$, whereas we need $a>0$. Without loss of generality, we will henceforth work with $k_\para > 0$. The first equation can be written as $\alpha( |\lambda|^2 + a) = -2k_\para$, such that $\alpha < 0$.

Combining the two conditions in Eq.~\eqref{eq:twoconditions}, we obtain
\begin{equation}
\alpha^3 + \alpha+ \frac{2k_\para}{|\lambda|^2} = 0. \label{eq:cubic}
\end{equation}
There are some interesting special cases to consider.

\textbf{Small momenta.} For $k_\para \approx 0$, we have $\alpha \approx 0$ such that we can drop the cubic term. We thus have $\alpha \approx -2k_\para/|\lambda|^2$. Remarkably, this implies that $\varepsilon_k = k_\para + \alpha |\lambda|^2 \approx  - k_\para$, i.e., the chirality has flipped! At the same time, the mode is spread out over a very large region: $a = \alpha^2 |\lambda|^2 \approx 4 k_\para^2/|\lambda|^2$ (remember that $\xi_\textrm{loc} = 1/a \approx |\lambda|^2/(4k_\para^2)$). Since $\lambda$ is a relevant perturbation, this localization length will blow up under RG.

\textbf{Intermediate momenta.} It is visually easy to see that there is a unique negative solution for $\alpha$ for any $k_\para > 0$ where the energy $\varepsilon_k = k_\para + \alpha|\lambda|^2$ vanishes. Indeed, this corresponds to $\alpha=-1$. This zero energy mode is exponentially localized with $a = |\lambda|^2$, i.e., $\xi_\textrm{loc} = 1/|\lambda|^2$. Remarkably, the RG-relevance of $\lambda$ tells us that $\xi_\textrm{loc} \to 0$ in the RG flow.

This new stable localized mode is at $k_\para = |\lambda|^2$. We can consider a small region around this, $k_\para = |\lambda|^2 + \delta k$. Similarly, $\alpha = -1 + \delta \alpha$. Eq.~\eqref{eq:cubic} then becomes $(1-\delta \alpha)^3 + (1-\delta \alpha) = 2 + 2\delta k / |\lambda|^2$, i.e., $2\delta \alpha \approx - \delta k / |\lambda|^2$. Hence, locally the dispersion is $\varepsilon_k = k_\para + \alpha|\lambda|^2 = \delta k + \delta \alpha |\lambda|^2 \approx \delta k/2$. I.e., the chirality is the same as the original mode, but the velocity is cut in half.

\textbf{Large momenta.} For large (positive) $k_\para$, we can ignore the linear term, i.e., $\alpha \approx - \sqrt[3]{2k_\para/|\lambda|^2}$. The dispersion relation is then $\varepsilon_k = k_\para ( 1+  \alpha|\lambda|^2/k_\para) \approx k_\para \left( 1 - \sqrt[3]{2} \times |\lambda|^{4/3} \times k_\para^{-2/3} \right) $.

\begin{figure}
\includegraphics[scale=1]{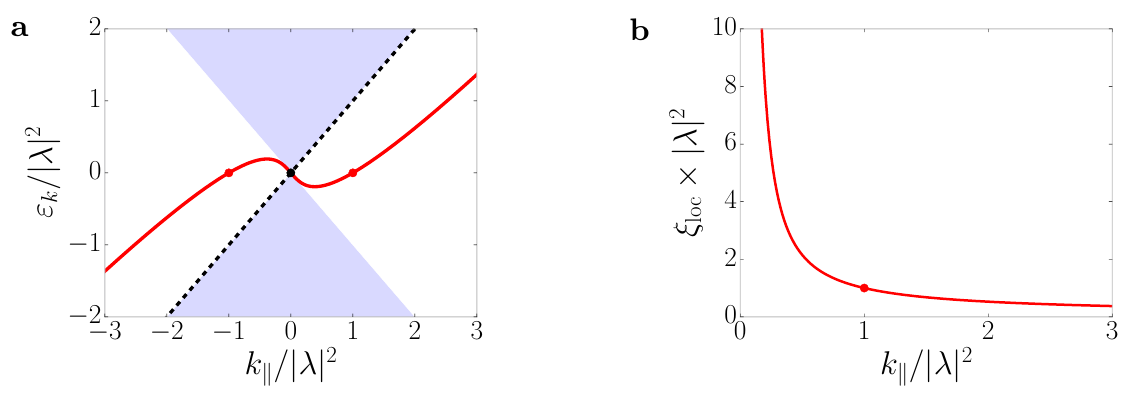}
\caption{
%Coupling a chiral edge mode to a delocalized bulk mode of the same chirality as described by Eq.~\eqref{eq:samechirality}.
\textbf{a}, A chiral edge mode (dashed line) is coupled to the Dirac cone (shaded region). More precisely, the coupling is to a bulk mode of the \emph{same chirality} as in Eq.~\eqref{eq:samechirality} (for \emph{opposite chirality}, see Fig.~\ref{fig:latticesamechirality}). The red curve is the renormalized dispersion. The red dots highlight the locations of the new zero-energy edge modes of the same chirality as the original mode. \textbf{b}, The localization length of the renormalized edge state. %shown in \textbf{a}.
The red dot corresponds to the zero-energy state(s). \label{fig:samechirality}}
\end{figure}

\textbf{Arbitrary momenta.} The closed expression is
\begin{equation}
\alpha = \beta - \frac{1}{3\beta} \qquad \textrm{with } \beta = \sqrt[3]{\sqrt{\kappa^2+\frac{1}{27}}-\kappa} \textrm{ where } \kappa = k_\para/|\lambda|^2.
\end{equation}
This then directly gives the closed expressions for the dispersion $\varepsilon_k = k_\para +\alpha|\lambda|^2$ and the localization length $\xi_\textrm{loc} = 1/a = \frac{1}{\alpha^2 |\lambda|^2}$. Note that it is natural to consider the dimensionless quantities $\varepsilon_k/|\lambda|^2$, $k_\para/|\lambda|^2$ and $\xi_\textrm{loc} |\lambda|^2$. These are plotted in Fig.~\ref{fig:samechirality}.

The above was for the special case that the (initial) edge mode had the same velocity as the gapless bulk modes. What if we replace the action of the edge mode with $\int v k_\para \eta^\dagger \eta$ (with $v>0$)? The same derivation goes through, with Eq.~\eqref{eq:condition} being replaced by
\begin{equation}
[\eta + \lambda \tilde \varphi + \alpha \lambda \tilde \psi, H_0 + V] = (v k_\para + \alpha |\lambda|^2) \eta - (k_\para + \alpha a ) \lambda \tilde \varphi + \left(k_\para + \frac{a}{\alpha}\right) \alpha \lambda \tilde \psi. 
\end{equation}
The dispersion is now $\varepsilon_k = v k_\para + \alpha |\lambda|^2$, with the conditions on $a$ and $\alpha$ being
\begin{equation}
(1+v)k_\para + \alpha |\lambda|^2 + \alpha a = 0 \qquad \textrm{and} \qquad a = \alpha^2 |\lambda|^2 + (v-1) \alpha k_\para = \alpha \left( \varepsilon_k - k_\para \right).
\end{equation}
It still follows that $\alpha$ is real and that its sign is opposite to that of $k_\para$. Combining the two conditions, we have
\begin{equation}
\alpha^3 + (v-1)\frac{k_\para}{|\lambda|^2} \alpha^2 + \alpha + (v+1) \frac{k_\para}{|\lambda|^2} = 0.
\end{equation}
As before, for $k_\para =0$, the only real solution is $\alpha =0$ (giving a delocalized mode). Let us now consider small but nonzero $k_\para$; we can then ignore the cubic and quadratic terms, such that $\alpha \approx -(v+1)k_\para/|\lambda|^2$. The dispersion is then $\varepsilon_k =v k_\para + \alpha|\lambda|^2 \approx - k_\para$. Remarkably, the dispersion now coincides with the bulk mode (of the opposite chirality)! As for the associated localization length,
\begin{equation}
\frac{1}{\xi_\textrm{loc}} = \alpha^2 |\lambda|^2 + (v-1)\alpha k_\para \approx \frac{(v+1)^2 k_\para^2}{|\lambda|^2} - \frac{(v^2-1) k_\para^2}{|\lambda|^2} = \frac{2(v+1)k_\para^2}{|\lambda|^2}.
\end{equation}
Hence, for any finite $k_\para \neq 0$ and $\lambda \neq 0$, we have an exponentially localized mode (as calculated for small $k_\para$). In fact, the stability applies even if $v$ is negative, as long as $v > -1 $. In other words, a right-moving mode is always stable, and a left-moving mode is stable as long as it doesn't lie inside the (blue) cone.

To determine the location of the zero-crossing, note that $\varepsilon_k=0$ is equivalent to $\frac{k_\para}{|\lambda|^2} = - \frac{\alpha}{v}$. Plugging this into the cubic equation:
\begin{equation}
\alpha^3 - (1-1/v) \alpha^3 + \alpha - (1+1/v) \alpha = 0.
\end{equation}
This readily simplifies to $\alpha^2-1=0$. The zero modes thus corresponds to $\alpha = \pm1$, as before. These are located at $k_\para = \pm \frac{|\lambda^2|}{v}$. This means that modes which were very fast (i.e., which lay deep in the blue/shaded/continuum area), are still very close to the bulk gapless mode. Relatedly, its localization length is larger: at the zero mode we have the general result that $\xi_\textrm{loc} = 1/|k_\para|$, so we have $\xi_\textrm{loc} = |v|/|\lambda|^2$.

\subsection{Conclusion}

We thus find that a chiral edge mode remains localized upon coupling it to a critical Dirac cone \emph{if} the nearby topological phase of the Dirac cone has the same chirality as the additional edge mode. The latter is encoded in the boundary condition of the Dirac cone.

However, there is an interesting edge reconstruction\cite{Halperin82,Chamon94,Barlas11} when we couple the edge mode to a bulk mode of the same chirality, as we saw in Fig.~\ref{fig:samechirality}. In particular, before coupling the two, we had a localized mode ($\psi_\textrm{loc}(k_\para) := \eta(k_\para)$) and a delocalized mode ($\psi_\textrm{deloc}(k_\para) := \int_0^\infty \psi(k_\para,x_\perp) \mathrm d x_\perp$). These are chiral modes with the same chirality, i.e., $[\psi_\textrm{(de)loc}(k_\para),H_0] = k_\para \psi_\textrm{(de)loc}(k_\para)$. (Note that for this half-infinite geometry we do \emph{not} have a delocalized mode of the other chirality due to the boundary condition $\varphi(x_\perp=0) = 0$.) After coupling the two, we found \emph{two} exponentially-localized edge modes with the chirality of the original mode (see red dots in Fig.~\ref{fig:samechirality}). Has our number of edge modes magically \emph{increased}? The catch is that we have changed the chirality of the delocalized mode! Indeed, this is also visually apparent in Fig.~\ref{fig:samechirality}, where the black dashed line has changed its chirality at $k_\para \approx 0$. Practically, this means that one can interpret the bulk Dirac cone now as a transition between $\mathcal C = -1 \leftrightarrow \mathcal C = 0$, rather than the original  $\mathcal C = 0 \leftrightarrow \mathcal C = 1$.

This intriguing outcome suggests that the bulk-boundary correspondence for topological invariants at criticality takes the form
\begin{equation}
\boxed{\quad \textrm{topological invariant} \quad = \quad \# \{ \textrm{localized edge modes} \} \quad+\quad \frac{1}{2} \times \# \{ \textrm{delocalized edge modes} \} \quad } \;, \label{eq:bulkboundary}
\end{equation} 
where the right-hand side has to be weighted by the proper chiralities/handedness/etc.
What we mean by the oxymoronic ``delocalized edge mode'' is that the delocalized mode is only well-defined in the presence of a particular boundary where it can terminate. E.g., above we saw that for the Dirac cone, the operator $\int_0^\infty \psi(k_\para,x_\perp) \mathrm dx_\perp$ was a chiral mode but $\int_0^\infty \varphi(k_\para,x_\perp) \mathrm dx_\perp$ was not.

What we have seen above is a nice illustration of Eq.~\eqref{eq:bulkboundary}. In particular, consider the Chern insulator transition between $\mathcal C= 1 $ and $\mathcal C =2$, which has $\mathcal C = \frac{3}{2}$ as discussed in the main text. One way of forming this is by taking a usual Dirac cone (i.e., the transition between $\mathcal C = 0 \leftrightarrow \mathcal C=1$) and a decoupled chiral edge mode. As discussed above, this decoupled system has one localized and one delocalized mode, satisfying Eq.~\eqref{eq:bulkboundary} in the form $\frac{3}{2} = 1+ \frac{1}{2}$. When we coupled the edge mode to a bulk mode of the opposite chirality, the situation did not qualitatively change. However, when we coupled it to the same chirality, we obtained two localized edge modes and switched the chirality of the delocalized mode. This still satisfies Eq.~\eqref{eq:bulkboundary} in the form $\frac{3}{2} = 2 + \frac{1}{2} \left( - 1 \right)$.

One might worry that the proposed bulk-boundary correspondence in Eq.~\eqref{eq:bulkboundary} does not give us a lower bound on the number of topologically-protected localized edge modes, since we can seemingly replace any number of edge modes into twice the number of delocalized modes. However, this is not the case: the number of delocalized modes is fixed by the bulk universality class, since they are a characteristic fingerprint of the bulk gapless cones. For instance, a Dirac cone only hosts one delocalized edge mode. This also nicely fits the narrative of our generalized index theorem (see Appendix~\ref{app:indextheorem}), where the bulk criticality was fixed by the number of purely-imaginary roots of a characteristic polynomial.

\section{Edge modes at critical Chern insulators: lattice perspective \label{app:Chernlattice}}
In this chapter, we consider a lattice model for the Chern insulator transitions. This will allow us to consider their experimentally-accessible local density of states and to test the field theory predictions of Appendix~\ref{app:Chernfield}.

\subsection{Periodic boundary conditions}
As already mentioned in the \hyperref[sec:methods]{Methods}, for any $\alpha \in \mathbb Z$ we define the following two-dimensional two-band model:
\begin{equation}
H_\alpha = \sum_{\bm k} \left( c_{\bm k,A}^\dagger, c_{\bm k,B}^\dagger \right) \mathcal H_\alpha(\bm k) \left( \begin{array}{c}
c_{\bm k,A} \\
c_{\bm k,B}
\end{array} \right) \qquad \textrm{with } \mathcal H_\alpha ( \bm k ) = \sin(\alpha k_x) \sigma^x - \sin(k_y) \sigma^y + (1-\cos(\alpha k_x) - \cos(k_y))\sigma^z. \label{eq:Halpha}
\end{equation}
We thus have $\bm h_\alpha(\bm k) = (\sin(\alpha k_x),-\sin(k_y),1-\cos(\alpha k_x) - \cos(k_y))$. Its spectrum is
\begin{equation}
\pm |\bm h_\alpha(\bm k)|= \pm \sqrt{ 3-2(\cos(\alpha k_x)+\cos(k_y)-\cos(\alpha k_x)\cos(k_y)) }.
\end{equation}
One can readily show that it is gapped (in fact, for $\alpha=0$ we have a flat band with $|h_0(\bm k)| = 1$; for $\alpha \neq 0$, we have $\min_{\bm k} |\bm h_\alpha(\bm k)| = 1$ and $\max_{\bm k}|\bm h_\alpha(\bm k)| = 3$). Moreover, these are Chern insulators with Chern number $\mathcal C = \alpha$:
\begin{equation}
\mathcal C = \frac{1}{4\pi} \iint \frac{\bm h_\alpha(\bm k)}{|\bm h_\alpha(\bm k)|^3} \cdot (\partial_{k_x} \bm h_\alpha(\bm k) \times \partial_{k_y} \bm h_\alpha(\bm k) ) \mathrm d k_x \mathrm d k_y = \frac{\alpha}{8\pi} \iint \frac{3-|\bm h_\alpha(\bm k)|^2}{|\bm h_\alpha(\bm k)|^3} \mathrm d k_x \mathrm dk_y = \alpha.
\end{equation}

\begin{figure}
	\includegraphics[scale=.9]{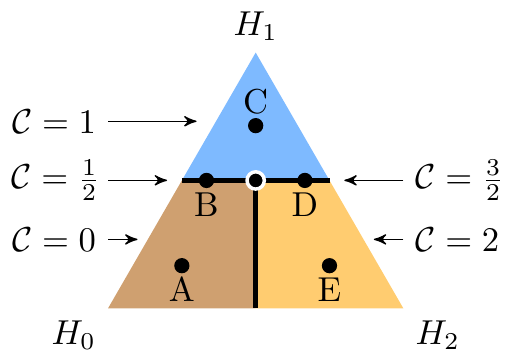}
	\caption{Phase diagram for the Chern insulator model in Eq.~\eqref{eq:Chernmodel}. The labels A through E denote the representative points defined in Eq.~\eqref{eq:points}. \label{fig:chernphasediagram}}
\end{figure}

We consider the interpolations between the Chern insulators with $\mathcal C=0,1,2$:
\begin{equation}
H = \frac{1}{a+b+c} \left( a H_0 + b H_1 + c H_2 \right) \qquad \textrm{with }a,b,c>0. \label{eq:Chernmodel}
\end{equation}
Its phase diagram is shown in Fig.~\ref{fig:chernphasediagram}. The direct transitions between $\mathcal C =0 \leftrightarrow \mathcal C=1$ and $\mathcal C=1 \leftrightarrow \mathcal C=2$ are topologically-distinct Dirac cones with $\mathcal C=\frac{1}{2}$ and $\mathcal C=\frac{3}{2}$, respectively. They are separated by a multi-critical point which, in this case, is described by a single cone which is linear in $k_y$ but quadratic in $k_x$.

To study the cases $\mathcal C=0,\frac{1}{2},1,\frac{3}{2},2$ in more detail, we choose a representative for each of these cases:
\begin{equation}
(a,b,c) = \left\{ \begin{array}{lll}
(4,1,1) & & \textrm{has } \mathcal C = 0, \\
(5,6,1) & & \textrm{has } \mathcal C = \frac{1}{2}, \\
(1,5,1) & & \textrm{has } \mathcal C = 1, \\
(1,6,5) & & \textrm{has } \mathcal C = \frac{3}{2}, \\
(1,1,4) & & \textrm{has } \mathcal C = 2.
\end{array} \right. \label{eq:points}
\end{equation}
These are indicated in Fig.~\ref{fig:chernphasediagram}.

\subsection{Open boundary conditions}

\subsubsection{Original model}
To study edge modes in these models, we need to express $H_\alpha$ (in Eq.~\eqref{eq:Halpha}) in real space variables $c_{\bm n,\lambda} = \frac{1}{\sqrt{L_x L_y}} \sum_{\bm k} e^{i \bm{k \cdot n}} c_{\bm k,\lambda}$ ($\lambda =A,B$):
\begin{align}
H_\alpha &= \frac{1}{2}\sum_{\bm n, \bm m} \left( c_{\bm n,A}^\dagger, c_{\bm n,B}^\dagger \right) T_{\bm n,\bm m}^{(\alpha)} \left( \begin{array}{c}
c_{\bm m,A} \\
c_{\bm m,B}
\end{array} \right) + h.c. \\
\textrm{with } 
T_{\bm n,\bm m}^{(\alpha)} &= \delta_{\bm n,\bm m} \sigma^z + \delta_{n_x,m_x} \delta_{n_y,m_y+ 1} (-i\sigma^y-\sigma^z) + \delta_{n_x,m_x+\alpha} \delta_{n_y,m_y} (i\sigma^x - \sigma^z) . \label{eq:realspace}
\end{align}
%System size is $2 L_x L_y$

\begin{figure}
	\includegraphics[scale=1]{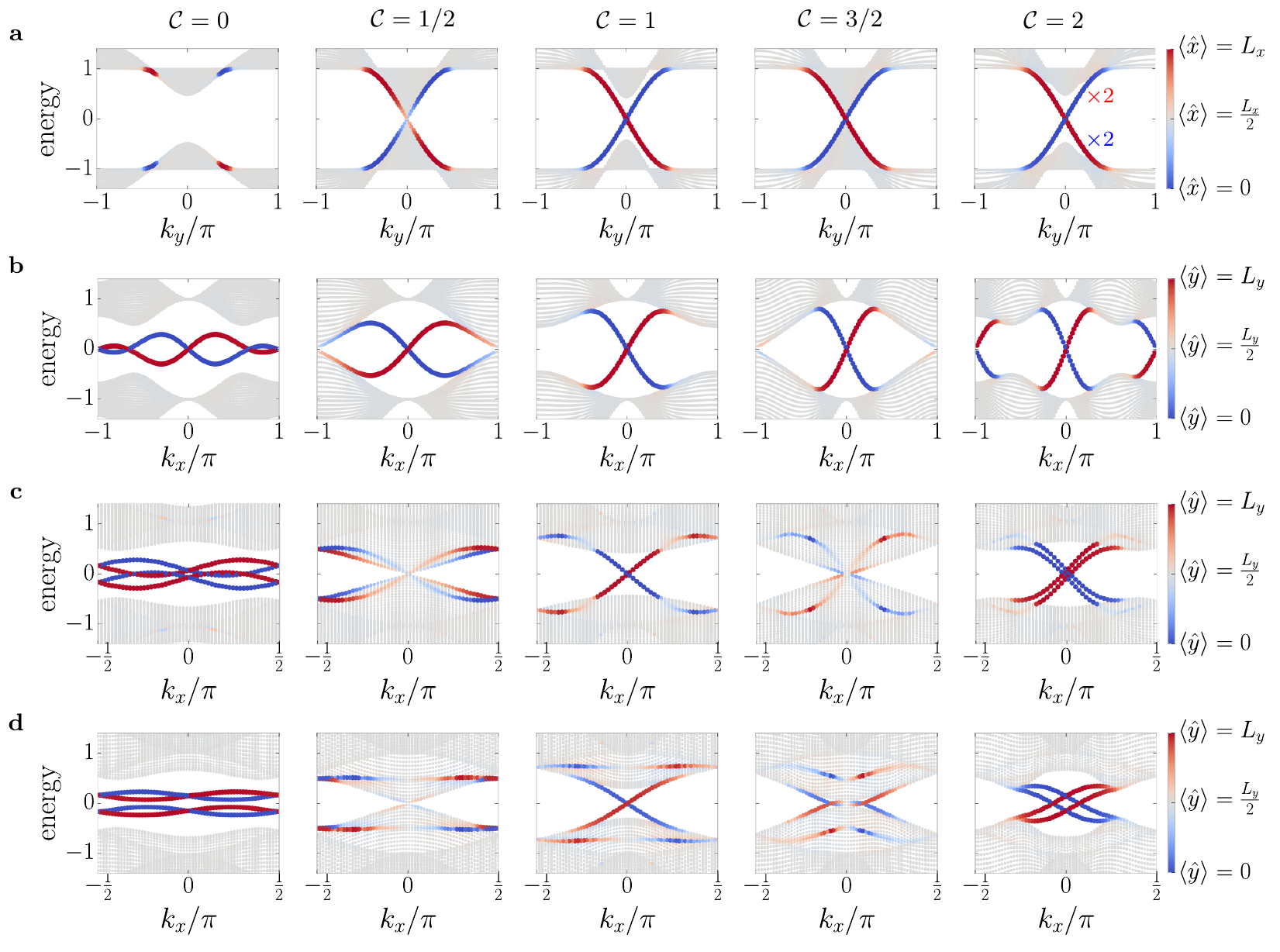}
	\caption{The energy spectrum for lattice models of gapped Chern insulators ($\mathcal C = 0,1,2$) and their transitions ($\mathcal C = \frac{1}{2},\frac{3}{2}$). One direction is periodic whereas the other has open boundary conditions. For each energy eigenstate, the color coding shows the mean position along the open direction.
	\textbf{a}, Open boundaries along the $x$-direction. We observe localized edge modes for $\mathcal C=\frac{3}{2}$ but not for $\mathcal C=\frac{1}{2}$.
	\textbf{b}, Open boundaries along the $y$-direction. We now even observe edge modes for $\mathcal C=\frac{1}{2}$, but these are only protected by momentum conservation (since the edge mode has $k_x =0$ but the bulk zero-energy modes have $k_x = \pi$).
	\textbf{c}, Since we are not interested in states protected by translation symmetry, we introduce a perturbation coupling $k_x = 0,\pi$. For this row, this coupling has strength $\lambda = 0.5$ (see text for details). This indeed kills the mode at $\mathcal C = \frac{1}{2}$. However, from this figure one might be tempted to conclude that the edge mode has also disappeared for $\mathcal C=\frac{3}{2}$.
	\textbf{d}, Ramping up the coupling to $\lambda =2$, we again see the edge mode for $\mathcal C=\frac{3}{2}$. Moreover, we quantitatively match this behavior to the field theory predictions of Appendix~\ref{app:Chernfield} which shows that the chiral edge mode (at $\mathcal C = \frac{3}{2}$) is there for any value of $\lambda$ (see Fig.~\ref{fig:comparison} and the text).
	\label{fig:chernlattice}}
\end{figure}

Let us first consider open boundaries along the $x$-direction. We can then still use $k_y$ as a good quantum number. Fig.~\ref{fig:chernlattice}(a) shows the energy levels obtained with free-fermion numerics for the representative points A through E (defined in Eq.~\eqref{eq:points}). Moreover, the color encodes the mean (real-space) position of each corresponding energy eigenstate. For the gapped Chern insulators with $\mathcal C=1$ and $\mathcal C=2$, we see the expected chiral edge states. For $\mathcal C=\frac{1}{2}$, we see that there are no edge modes at criticality as evidenced by the color coding fading out near the gapless point (corresponding to delocalized bulk states). However, for $\mathcal C=\frac{3}{2}$, we see a chiral edge state. Note that we cannot blame the edge mode's stability on the conservation of $k_y$ since the bulk Dirac cone and the edge state occur at the same value $k_y = 0$. This is the same set-up which we used to calculate the local DOS as already detailed in the \hyperref[sec:methods]{Methods} and as shown in Fig.~\ref{fig:exp} of the main text.

It is instructive to also consider boundaries along the $y$-direction (keeping periodicity along the $x$-direction). This is plotted in Fig.~\ref{fig:chernlattice}(b). Interestingly, in this case, even the trivial phase has edge states; however, these are not protected (indeed, the net chirality of the edge states sums to zero \emph{for each edge}). Similarly, both critical points $\mathcal C=\frac{1}{2}$ and $\mathcal C=\frac{3}{2}$ display edge modes; we claim that these are unprotected in the former case but stable in the latter case. More precisely, if we would add a perturbation that allows $k_x \approx 0$ to scatter into $k_x \approx \pi$, we expect that the edge mode disappears for $\mathcal C = \frac{1}{2}$ but persists for $\mathcal C = \frac{3}{2}$. To test this hypothesis, we need to enlarge our unit cell.

\subsubsection{Enlarged unit cell}

To enlarge the unit cell along the $x$-direction, we replace the labeling $n_x$ by $n_x = 2 \tilde n_x + a$, i.e., $n_x = (\tilde n_x,a)$ with $a=0,1$. Similarly, $m_x = 2 \tilde m_x + b$. Above, we used the Pauli matrices $\bm \sigma$ for our two-band model in Eq.~\eqref{eq:realspace}. We now consider the above label $a=0,1$ as an additional internal index, denoting the corresponding Pauli matrices as $\bm \tau$. In this rewriting, we now have a four-band model.

More precisely, if $\alpha = 2\tilde \alpha$ is even, then the real-space hopping matrix in Eq.~\eqref{eq:realspace} can be expressed as
\begin{equation}
T_{\bm n,\bm m}^{(\alpha)} = \delta_{\bm n,\bm m} \sigma^z \tau^0 + \delta_{\tilde n_x,\tilde m_x} \delta_{n_y,m_y+ 1} (-i\sigma^y-\sigma^z) \tau^0 + \delta_{\tilde n_x,\tilde m_x+\tilde \alpha} \delta_{n_y,m_y} (i\sigma^x - \sigma^z) \tau^0.
\end{equation}
If $\alpha = 2 \tilde \alpha + 1$ is odd, we use the fact that 
\begin{equation}
\delta_{n_x,m_x+\alpha} = \delta_{2\tilde n_x+a,2(\tilde m_x+\tilde \alpha)+b+1} = \delta_{\tilde n_x,\tilde m_x + \tilde \alpha} \frac{\tau^x - i \tau^y}{2} + \delta_{\tilde n_x, \tilde m_x + \tilde \alpha + 1} \frac{\tau^x + i \tau^y}{2}.
\end{equation}
In the special case of $\alpha=1$ (i.e., $\tilde \alpha = 0$) we thus obtain
\begin{equation}
T_{\bm n,\bm m}^{(1)} = \delta_{\bm n,\bm m} \left( \sigma^z\tau^0 + (i\sigma^x - \sigma^z) \frac{\tau^x - i \tau^y}{2} \right) + \delta_{\tilde n_x,\tilde m_x} \delta_{n_y,m_y+ 1} (-i\sigma^y-\sigma^z) \tau^0 +\delta_{\tilde n_x,\tilde m_x+1} \delta_{n_y,m_y} (i\sigma^x - \sigma^z) \frac{\tau^x + i \tau^y}{2}.
\end{equation}

Thus far, this is merely a rewriting of the same model. The main purpose was to halve the Brillouin zone along the $x$-direction. More precisely, we can effectively replace $\delta_{\tilde n_x,\tilde m_x+\tilde \alpha} \to e^{-i\tilde k_x \tilde \alpha}$, where we can associate $\tilde k_x = 2 k_x$. In other words, the original values $k_x =0$ and $k_x = \pi$ now both correspond to $\tilde k_x = 0 \mod 2\pi$. Of course, as long as we do not perturb the Hamiltonian, there is a hidden conservation law that prevents the two from coupling. However, this will be broken by a generic coupling that acts non-trivially on the $\tau$-variables. In particular, we find it convenient to consider the perturbation $\lambda i \delta_{\tilde n_x,\tilde m_x} \delta_{n_y,m_y+ 1} \sigma^x \tau^z + h.c.$, since this does not seem to change the overall structure of the phase diagram (i.e., we automatically stay tuned to criticality).

If we now turn on this perturbation with a coupling strength $\lambda=0.5$, then Fig.~\ref{fig:chernlattice}(b) is replaced by Fig.~\ref{fig:chernlattice}(c)---note that the $x$-axis has half the original range due to halving the Brillouin zone. We see that for $\mathcal C = \frac{1}{2}$, the edge mode has successfully coupled to the bulk Dirac cone and has disappeared. However, the edge mode also seems to have disappeared for $\mathcal C=\frac{3}{2}$. This is not true: it is there, but it turns out to not be readily visible at this resolution---as we clarify now.

To confirm the stability of the edge mode at $\mathcal C= \frac{3}{2}$, let us ramp up the perturbation strength to $\lambda=2$. The result is plotted in Fig.~\ref{fig:chernlattice}(d). The chiral edge mode is now indeed visible. In fact, it has the same qualitative structure as the edge reconstruction we saw in Appendix~\ref{app:Chernfield}. There, we found that coupling a chiral edge mode to a delocalized bulk state of the same chirality led to the dispersion in Fig.~\ref{fig:samechirality}(a). This is very similar to the dispersion in Fig.~\ref{fig:chernlattice}(d) for $\mathcal C=\frac{3}{2}$.

\begin{figure}
	\includegraphics[scale=1]{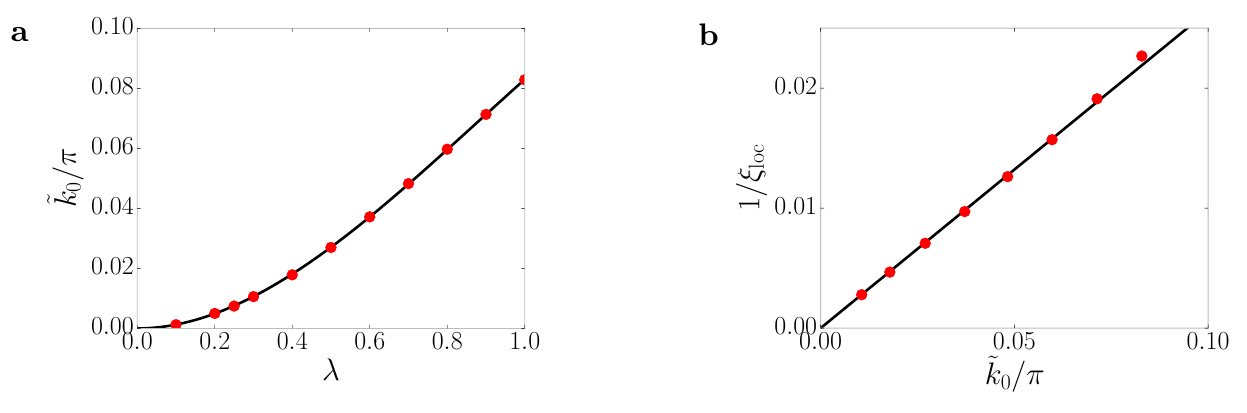}
	\caption{Confirming the field theory predictions from section~\ref{sec:field} of Appendix~\ref{app:Chernfield} (black line) in a lattice model (red dots). Here, $\tilde k_0$ denotes the edge mode's momentum (at zero energy) along the boundary in the reduced/halved zone (see text; there is a factor of two difference from the original momentum $k_x =2\tilde k_x$) and $\lambda$ is the perturbation strength. The edge mode's localization length is denoted as $\xi_\textrm{loc}$. The cases $\lambda=0.5$ and $\lambda=2$ are studied in more detail in Fig.~\ref{fig:chernlattice}(c) and (d), respectively. \label{fig:comparison}}
\end{figure}

This relation to the field theory prediction in Appendix~\ref{app:Chernfield} is not just an accidental similarity. We confirm that this is a quantitative match. More precisely, in Appendix~\ref{app:Chernfield} we derived that the edge momentum depends on the square of the perturbation strength. We confirm this for the lattice model in Fig.~\ref{fig:comparison}(a). Moreover, we also confirm the field theoretical prediction that the edge mode's inverse localization length is also proportional to the square of the perturbation strength---or by the aforementioned---proportional to the edge momentum. This is confirmed in Fig.~\ref{fig:comparison}(b).

This non-trivial confirmation of the field theory predictions of Appendix~\ref{app:Chernfield} also implies that the chiral edge mode (for $\mathcal C=\frac{3}{2}$) is there for any arbitrarily small value of the perturbation $\lambda$. This is indeed supported by Fig.~\ref{fig:comparison}.

\section{Deriving boundary conditions from spatial interfaces \label{app:bc}}

Here, our goal is to show how to derive the boundary condition of a fermionic field theory. The only required information is: (1) the bulk Hamiltonian, and (2) which gapped phase is the trivial phase. We illustrate this here for a one-dimensional Hamiltonian of the form $H = i \int \tilde \chi(x) D_x \chi(x) \mathrm dx $ where $D_x$ is an arbitrary differential operator. For concreteness, we take these fields to be Majorana fields, i.e., $\{ \tilde \chi(x),\chi(y) \} = 0$ and $\{ \chi(x),\chi(y)\} = 2\delta(x-y) = \{ \tilde \chi(x), \tilde \chi(y)\} $, although this is not essential to this approach.

The idea is as follows: we will start from an infinite system and consider a spatial interface where $x<0$ is in the trivial phase (and $x>0$ is in principle arbitrary). This is a soft boundary. In the limit of taking the region $x<0$ to its RG fixed point limit, we will effectively have a half-infinite system $x \geq 0$ with a hard boundary at $x=0$. We will determine its boundary condition by ensuring that any zero-energy mode of the original interface is preserved in this hard-boundary-limit. More precisely, suppose that the original interface had a mode of the form $\chi_\textrm{loc} = \int_{-\infty}^{x_0} f(x) \chi(x) \mathrm d x$ (with $x_0\geq 0$) which is a zero-energy mode aside from its endpoint, i.e., $[\chi_\textrm{loc},H] = \mathcal O(x_0)$ (where $\mathcal O(x)$ is a local operator near $x$). Moreover, suppose that in the aforementioned hard-boundary-limit, $f(x) \to 0$ for $x<0$. We then determine the boundary condition by ensuring that the new (truncated) operator $\chi_\textrm{loc} \to \int_0^{x_0} f(x) \chi(x) \mathrm d x$ still obeys $[\chi_\textrm{loc},H] = \mathcal O(x_0)$. It is not hard to see that in practice, it is sufficient to take $x_0=0$, in which case the boundary condition is $\mathcal O(x=0) \equiv 0$.

\subsection{The general case}

Before illustrating this general approach in more detail, let us first note the following useful result: 
\begin{equation}
\boxed{ \left[\int_{-\infty}^0 f(x) \chi(x) \mathrm d x, \frac{1}{2} \int_{-\infty}^{\infty} \tilde \chi(x) D \chi(x) \mathrm d x \right] =  \sum_{i=0}^\infty  \left( \sum_{m=0}^\infty a_{m+i+1} (\partial^m f)(0) \right) (-\partial)^i\tilde \chi(0) - \int_{-\infty}^0 \left( D f \right)(x) \tilde \chi(x) \mathrm dx } \label{eq:bc}
\end{equation}
where $D = \sum_{n=0}^\infty a_n \partial^n$ and presuming that $\lim_{x \to - \infty}f(x) = 0$.

\begin{proof}
	First, note that the left-hand side of Eq.~\eqref{eq:bc} can be simplified (where $D^\dagger$ is shorthand for $\sum_n a_n (-\partial)^n$):
	\begin{equation}
	\left[ \int_{-\infty}^0 f(x) \chi(x) \mathrm d x, -\frac{1}{2} \int_{-\infty}^{\infty} \chi(x) D^\dagger \tilde \chi(x) \mathrm d x \right] = -\frac{1}{2} \int_{-\infty}^{0} \mathrm d x \int_{-\infty}^\infty \mathrm d y \; f(x) \underbrace{\{ \chi(x),\chi(y) \}}_{=2\delta(x-y)} D^\dagger \tilde \chi(y) \mathrm d y = -\int_{-\infty}^{0} f(x) D^\dagger \tilde \chi(x) \mathrm d x.
	\end{equation}
	Secondly, by the linearity of Eq.~\eqref{eq:bc}, it is sufficient to prove it for $D = \partial^n$, which we do by induction. For $n=0$, this is true by inspection. For $n>0$, the simplified claim is
	\begin{equation}
	\int_{-\infty}^{0} f(x) (-\partial)^n \tilde \chi(x) \mathrm d x = \int_{-\infty}^0 (\partial^n f)(x) \tilde \chi(x) \mathrm dx - \sum_{m=0}^{n-1} (\partial^m f)(0) (-\partial)^{n-m-1}\tilde \chi(0). \label{eq:simplified}
	\end{equation}
	Let us presume that Eq.~\eqref{eq:simplified} holds for $n=N-1$. We now prove it for $n=N$:
	\begin{align}
	\int_{-\infty}^{0} f(x) (-\partial)^{N} \tilde \chi(x) \mathrm d x
	& = - f(0) (-\partial)^{N-1} \tilde \chi(0) + \int_{-\infty}^0 f' (x) (-\partial)^{N-1} \tilde \chi(x) \mathrm d x \\
	&= - f(0) (-\partial)^{N-1} \tilde \chi(0) + \left( \int_{-\infty}^0 (\partial^{N-1} f')(x) \tilde \chi(x) \mathrm dx - \sum_{m=0}^{N-2} (\partial^m f')(0) (-\partial)^{N-m-2}\tilde \chi(0) \right).
	\end{align}
	After a change of index ($\tilde m := m+1$), this indeed agrees with Eq.~\eqref{eq:simplified} for $n=N$.
\end{proof}

For the above to be useful, we need solutions $(Df)(x) = 0$ (such that the last term in Eq.~\eqref{eq:bc} vanishes), and such that $f(x) \to 0$ when the outside `trivial' region flows to its RG fixed point (such that the LHS of Eq.~\eqref{eq:bc} vanishes). In that case, it gives us boundary conditions through $\sum_{i=0}^\infty  \left( \sum_{m=0}^\infty a_{m+i+1} (\partial^m f)(0) \right) (-\partial)^i\tilde \chi(0) = 0$. Similarly, by starting from $\int_{-\infty}^{0} g(x) \tilde \chi(x) \mathrm d x$, we get that any $g(x)$ such that $(D^\dagger g)(x)=0$ gives us a boundary condition $\sum_{i=0}^\infty  \left( \sum_{m=0}^\infty a_{m+i+1} (\partial^m g)(0) \right) (-\partial)^i \chi(0) = 0$.

\subsection{Examples}

\subsubsection{One-parameter family}

Let us consider $D = \partial + m$.  We look for functions annihilated by $D$, given by $f(x) = e^{-mx}$. If the trivial phase corresponds to $m<0$, then we indeed have that $f(x) \to 0$ if the trivial region $x<0$ is brought to its fixed point limit ($m \to -\infty$). By the above, we thus derive the boundary condition $2i f(0) \tilde \chi(0) = 2i \tilde \chi(0) = 0$. However, if $m>0$ is the trivial phase, then $f(x)$ blows up in the RG limit. This is a sign that we should instead look at functions $g(x)$ annihilated by $D^\dagger$, i.e., $g(x) = e^{mx}$. By the above, we now obtain the boundary condition $\chi(0) = 0$. In summary,
\begin{equation}
\textrm{if } H = i\int_0^{\infty} \tilde \chi(x) (\partial_x + m) \chi(x) \mathrm dx \textrm{ is } \left\{ \begin{array}{ccc}
\textrm{ trivial for } m < 0 \; & \Rightarrow &\textrm{ we have the boundary condition } \tilde \chi(0) = 0, \\
\textrm{ trivial for } m > 0 \; & \Rightarrow &\textrm{ we have the boundary condition } \chi(0) = 0.
\end{array} \right.
\end{equation}
This agrees with our discussion in the \hyperref[sec:methods]{Methods}.

\subsubsection{Two-parameter family}

We now consider $D = -\partial^2 - \kappa \partial + m$ as in Eq.~\eqref{eq:criticalinterface} of the main text. Let us determine its nearby phase diagram, driven by the two relevant parameters (with scaling dimensions $[\kappa] = 1$ and $[m]=2$). This is straightforward since we can read off the single-particle energy spectrum: $\varepsilon_k = | k^2 - i \kappa k + m| = \sqrt{(k^2+m)^2 +\kappa^2 k^2} $. This is gapped if $m> 0$. However, if $m < 0$, it is gapless at $k = \pm \sqrt{|m|}$ (giving a conformal field theory with central charge $c=1$), unless $\kappa \neq 0$. If $m=0$ and $\kappa \neq 0$, we have a single massless relativistic Majorana mode (central charge $c=1/2$). The nearby phase diagram thus looks as follows:
\begin{center}
	\includegraphics[scale=0.9]{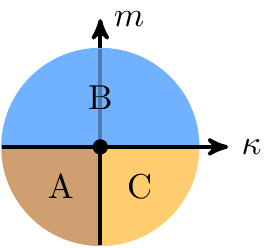}
\end{center}
One can derive the following boundary conditions:
\begin{equation}
\left\{ \begin{array}{ccccccc}
\textrm{if A is trivial} \qquad &\Rightarrow& \qquad \tilde \chi(0) & = & \partial \tilde \chi(0) & = & 0, \\
\textrm{if B is trivial} \qquad &\Rightarrow& \qquad \tilde \chi(0)& =& \chi(0)& =& 0,  \\
\textrm{if C is trivial} \qquad &\Rightarrow& \qquad \chi(0)& = & \partial \chi(0) &=& 0.
\end{array} \right.
\end{equation}
To see this, note that the functions annihilated by $D$ are $f_\pm(x) = e^{a_\pm x}$ with $a_\pm = -\frac{\kappa}{2} \pm \sqrt{\left(\frac{\kappa}{2}\right)^2 + m}$. Similarly, the functions annihilated by $D^\dagger$ are $g_\pm(x) = e^{b_\pm x}$ with $b_\pm = \frac{\kappa}{2} \pm \sqrt{\left(\frac{\kappa}{2}\right)^2 + m}$.

If A is trivial, then in the region $x<0$ we have $m,\kappa < 0$. In that case, both choices $a_\pm$ are positive, giving us a localized function that flows to zero in the RG fixed point limit $m,\kappa \to - \infty$. By the above, we thus derive the conditions
\begin{equation}
\left(-\kappa f_\pm(0) - f_\pm'(0)\right)\tilde \chi(0) + f_\pm(0) \partial \tilde \chi(0) = 0 \quad \Rightarrow \quad
\left(-\kappa - a_\pm \right)\tilde \chi(0) +  \partial \tilde \chi(0) = 0. \label{eq:bcondition}
\end{equation}
More precisely, this only holds in the limit $m,\kappa \to - \infty$. We see that in this case, this effectively forces $\tilde \chi(0) = \partial \tilde \chi(0) \to 0$.

If B is trivial, then the region $x<0$ has $m>0$ (and $\kappa$ is arbitrary). In this case, only one choice of $a_\pm$ gives a decaying mode. Let us say $a_+$ for concreteness. As in Eq.~\eqref{eq:bcondition}, we then obtain $(-\kappa-a_+) \tilde \chi(0) + \partial \tilde \chi(0) = 0$, i.e., $\tilde \chi(0) \to \frac{\partial \tilde \chi(0)}{\kappa + a_+}$. In the RG limit of $x<0$, this pins $\tilde \chi(0) = 0$ but it does not pin $\partial \tilde \chi(0)$. Similarly, one of the $b_\pm$ gives a localized mode, pinning $\chi(0) = 0$.

If C is trivial, we can reduce it to the first scenario by swapping $\chi \leftrightarrow \tilde \chi$.

\end{document}